\documentstyle[11pt]{article}

\begin{document}
\oddsidemargin .3in
\evensidemargin 0 true pt
\topmargin -.4in


\def\ra{{\rightarrow}}
\def\a{{\alpha}}
\def\b{{\beta}}
\def\l{{\lambda}}
\def\eps{{\epsilon}}
\def\T{{\Theta}}
\def\t{{\theta}}
\def\co{{\cal O}}
\def\car{{\cal R}}
\def\caf{{\cal F}}
\def\cs{{\Theta_S}}
\def\pr{{\partial}}
\def\tri{{\triangle}}
\def\na{{\nabla }}
\def\S{{\Sigma}}
\def\s{{\sigma}}
\def\sp{\vspace{.15in}}
\def\hs{\hspace{.25in}}

\newcommand{\be}{\begin{equation}} \newcommand{\ee}{\end{equation}}
\newcommand{\bea}{\begin{eqnarray}}\newcommand{\eea}
{\end{eqnarray}}


\begin{titlepage}
\topmargin= -.2in
\textheight 9.5in

\baselineskip= 17 truept

\vspace{.3in}

\noindent
{\LARGE\bf Electromagnetic Strings: }

\vspace{.05in}

\noindent
{\LARGE\bf Complementarity between Time and Temperature}

\vspace{.4in}

\noindent
{{{\Large Supriya Kar}\footnote{supriya@iitk.ac.in}}
{{\Large and Sudhakar Panda}\footnote{panda@mri.ernet.in}}}

\vspace{.2in}

\noindent
{\large ${}^1$Department of Physics\\
Indian Institute of Technology,
Kanpur 208 016, India}

\vspace{.2in}

\noindent
{\large ${}^2$Harish-Chandra Research Institute\\
Chhatnag Road, Jhusi,
Allahabad 211 019, India}

\vspace{.1in}


\thispagestyle{empty}

\vspace{.6in}
\begin{center}
{\bf Abstract}
\end{center}

\vspace{.2in}


\baselineskip= 14 truept

We investigate some of the intricate features in a gravity decoupling limit of
a open bosonic string theory, in a constant electromagnetic (EM-) field. 
We explain the subtle nature of space-time
at short distances, due to its entanglement with gauge field windings in the theory.
Incorporating the mass-shell condition,
we show that the time coordinate is small, of the order 
of EM-string scale, and the space coordinates are large.
We perform a careful analysis in the critical regime
to describe the decoupling of a series of gauge-string windings in 
successions, just below the Hagedorn temperature. We argue for the condensation 
of gauge-string at the Hagedorn temperature, which is followed by the decoupling
of tachyonic particles. We demonstrate the phenomena by
revoking the effective noncommutative dynamics for $D(3)$-brane 
and obtain nonlinear corrections to $U(1)$ gauge theory. 
We discuss the spontaneous breaking of noncommutative $U(1)$ symmetry 
and show that the Hagedorn phase is described by the noninteracting gauge particles.
The notion of time reappears in the phase at the expense of temperature.
It suggests a complementarity between two distinct notions, time and
temperature, at short distances.

\vspace{.3in}




\end{titlepage}

\baselineskip= 18 truept

\section{Introduction}

\hs
Conventional notion of space-time becomes subtle, at short distances,
due to the existence of intrinsic bounds in string theory. 
There, the space-time windings grow exponentially and may invalidate
the underlying Riemannian manifold. Thus, it requires a new setting to describe the
physical phenomena at short distances.

\sp
For instance, an equilibrium thermal string description 
encounters a bound due to the Hagedorn temperature
in the theory. Apart from this, there are other stringy bounds in the theory.
Recent idea of Dirichlet (D-) branes in string theory favours the assertion and
allows one to probe shorter distances than the string scale \cite{douglas}. 
The presence of
electromagnetic (EM-) field in string theory \cite{fradkin,ferrer}
further provides evidence for a bound  
and allows one to access even a smaller 
distance. However, the minimal length scale in the theory 
is bounded{\footnote{A bound on  electric field, eliminates
the tachyons, otherwise present in the (super) string theory.}}
by a critical value of electric ($E$-) field \cite{odintsov}-\cite{gukov}.
Under most of the
circumstances{\footnote{Except for the perpendicular $E$- and
$B$-fields.}}, the magnetic ($B$-) field can also be
seen to be bounded by the value of the critical $E$-field in the theory.
On one hand, an $E$-field introduces the notion of temperature and on the other,
it initiates small distance phenomena, leading to space-time noncommutativity in
string dynamics \cite{connes}. In this work, we focus on these two alternate
descriptions, in a open bosonic string theory and find a complementarity between
time and temperature. For a comprehensive review on
the aspects of strings in EM-field, we refer the reader to ref.\cite{ambjorn}.

\sp
In the recent past, the notion of noncommutative space-time has been established in 
a open string theory, at its boundary, due to the presence of
a constant two-form $b_{\mu\nu}$ in the theory. Since the global
mode $b_{\mu\nu}$ cannot be gauged away by any gauge transformations in
the theory, the
noncommutativity among the space-time coordinates becomes 
inevitable there.
For instance, a strong
$B$-field in absence of an $E$-field decouples 
all the stringy modes from the theory \cite{jabbari1}-\cite{chu1}. In an effective
description, it corresponds to a non-commutative $U(1)$ theory, 
which is perturbatively unitary \cite{minwalla}. However in presence of a 
non-zero $E$-field and vanishing $B$-field, some of the light stringy
modes do not decouple from the theory and in the effective description, it
corresponds to a noncommutative open string theory (NCOS), 
instead of a field theory \cite{susskind}-\cite{barbon1}.
In the context, it is 
been realized that theories with temporal noncommutativity are
consistent quantum mechanically. Nevertheless,
it is certain that the space-time
noncommutative field theories do not possess a unitary S-matrix and
lead to acausal behaviour. It means, a stable
field theoretic limit does not exist in a space-time noncommutative string
theory. In other words, the low energy domain is described by a critical $E$-field,
which can be seen to  be necessary to get rid of the instabilities even at the
classical level.{\footnote{In a quantum picture,
the electric field generates charged string pairs
from the vacuum, which leads to enormous energy and hence
instability in the theory \cite{bachas}.}}. 

\sp
In an alternate, thermal, description for the effective NCOS, a bound on $E$-field
can be seen to introduce Hagedorn temperature. In fact,
the notion of finite temperature becomes relevant as soon as an $E$-field is introduced
in the theory \cite{ferrer}. At this point, it is inspiring to recall the analogy
between Hagedorn temperature in
string theory and the deconfining transition in QCD \cite{polyakov,susskind2}.
As a consequence,
the study of high temperature string phase is believed to describe a spontaneously
broken non-abelian gauge symmetry, which in turn leads to deconfinement in
QCD \cite{gross}. On the other hand, 
in a bosonic string theory, it was shown that some of the winding modes
become massless and describes a phase transition at the Hagedorn temperature \cite{sathia}.
Subsequently, a first order phase transition has been confirmed
at Hagedorn temperature, where the mass-less mode behaves as
the order parameter in the theory \cite{atick}. Thus, 
the Hagedorn behaviour is intertwined with NCOS, due to the inherent notion of temperature
in the theory.
Recent investigations \cite{mukund}-\cite{abel} for Hagedorn transition in NCOS theory has 
been shown to be second order, due to the absence of massless closed string modes. 

\sp
In this paper, we investigate some of the features  
entangled with the global modes of a two-form 
$b$-field,{\footnote{We recall that an arbitrary $b$-field is an
integral part of the massless closed string backgrounds. It combines
with the $U(1)$ field strength $F= dA$, present at the string boundary,
to retain the gauge invariance in the theory.}} 
in a open bosonic string theory, at its gravity decoupling limit \cite{gopa1}.
We begin with a ($3+1$)-dimensional
space-time, describing a $D(3)$-brane at the string world-sheet boundary{\footnote{An 
arbitrary dimensional space-time 
generalization is straightforward. For $N$ number of $D(3)$-branes, the gauge group
at the string boundary becomes $U(N)$.}}.
The string mode expansion 
in presence of a constant EM-field (${2\pi\a' F} + b$) is reviewed.
We perform a careful analysis for the light windings, after decoupling all the
massless closed string modes from the theory. 
In the limit $E\ra 
E_c$ and $\a' \ra 0$, the massless closed string modes decouple 
from the open string and leaves behind a
nongravitational, "rigid" 
electric string with string tension $T\ra \infty$.{\footnote{$E$-field
in the theory introduces a 
stretch parallel to the string, where as a
$B$-field gives a stretch
along transverse direction to propagation. A priori, a 
stretch due to $B$-field gives rise to a width, unlike to tension due to the
$E$-field, since a $B$-field does no work. However, 
a $B$-field is a source for space-like
windings and contributes to energy in the theory.}}
Alternately, in an  effective description,
the theory is known to possess noncommutativity among its space and time 
coordinates and becomes tension-less $T_{\rm eff}\ \ra 0$ \cite{gukov}.
At a critical electric field ($E_c$), some of the light mass string states remain in the
effective NCOS theory and describes the dynamics for "long" strings there.
When a $B$-field is introduced parallel to $E$-field, the bound on the $E$-field
becomes a bound on $B$-field as well. In this case, an $E$-string may be 
generalized to the EM-string. Since under a strong-week coupling duality, the fields
get interchange $E\leftrightarrow B$, the relative field strength becomes unimportant
for the case $E\parallel B$, which is unlike to that in $E\perp B$.
It implies that the orientation of $B$-field with respect to $E$-field plays an
important role in the gravity decoupling limit.

\sp
On the other hand, the
difference between EM-string theory and that of the fundamental
string is in their respective length scales associated with 
independent sources in the theory. 
Needless to mention that the prime source to the fundamental string scale is gravity and
that to an effective noncommutative string is the light mass gauge field.
In addition, the noncommutative string is defined with a minimal length scale, very
unlike to that in a fundamental string.
It leads to a conceptual difference between the
notion of uncertainty among the space-time coordinates
in string theory \cite{yoneya} and in a noncommutative string theory 
\cite{connes}-\cite{barbon1}.
The minimal noncommutative string scale does not allow an independent field theoretic 
description in the theory, instead describes a NCOS theory. It implies that a
space-time noncommutative field theory may not be obtained from the NCOS theory.
Nevertheless, a field theoretic domain is consistent within a NCOS theory. 
In this context, we analyze the NCOS modes to explain the cancellation of
unphysical states present, otherwise, in a field theoretic domain.

\sp
In this paper, we encounter a mixing of time with space components from the
nonzero momentum modes in the theory.
Incorporating the mass-shell condition, we
show that the time-like dimension is small
and the space-like dimensions are large in a ($3+1$)-dimensional 
EM-string theory. The tiny 
time-like modes give rise to a time translation symmetry and introduces
the notion of finite temperature in the EM-string theory. 
We consider the phase transitions associated with the decoupling of massless
and some of the light mass string modes and obtain an expression for
the Hagedorn temperature in the theory. We argue for the condensation of EM-string
at Hagedorn temperature. Subsequently, the light modes become
tachyonic and decouple from the theory to describe the Hagedorn phase.
We present a brief analysis on
a space-time noncommutative $U(1)$ theory, in the critical regime. The spontaneous
breaking of noncommutative $U(1)$ symmetry \cite{seiberg,campbell} is discussed
in the Hagedorn phase to comment on the complementarity between the notion of time
and temperature in the theory.

\sp
The paper is organized as follows. In section 2.1, we describe the $D(3)$-brane
dynamics in presence of a constant EM-field to make our presentation
self-contained. In section 2.2, we obtain a bound on $E$-field, by briefly
reviewing the $D(3)$-brane dynamics described by the Dirac-Born-Infeld
(DBI) action and its noncommutative counterpart in the effective description.
In section 2.3, we demonstrate the
importance of string  modes to unitarity in a space-time noncommutative theory
and obtain the Hamiltonian for the propagation of NCOS. We
describe in some detail on the nature of time and space coordinates in section
3.1 for $E\perp B$ and in section 3.2 for $E\parallel B$. In section 3.3, we
confirm the mixing of time with space components from the nonzero Fourier modes
in the theory and exhibit the left-right symmetry in the gravity decoupled NCOS  
(EM-string) theory. 
In section 4.1, we describe the thermal modes and obtain an expression for
the Hagedorn temperature in the theory. In section 4.2, we describe phase
transitions,
just below the Hagedorn temperature and account for a time-like 
noncommutative $U(1)$ theory. In section 4.3, we elaborate on the Hagedorn phase
in the context and its similarities with the deconfining phase in QCD. We
summarize our results and outline on some of the further directions in section 5.

\section{Strings in constant electro-magnetic field}

\subsection{Set up: D(3)-brane dynamics}

\hs
Consider a $D(3)$-brane in 
a open bosonic string  theory. The $D(3)$-brane world-volume
dynamics can essentially be described by a open string 
at its boundary. Since a $D(3)$-brane describes a (3+1)-dimensional
space-time $X^{\mu}$ (${\mu = 0,1,2,3}$) embedded in string theory, 
the transverse space to the
brane is described by Dirichlet boundary conditions $\delta X^{\perp} = 0$.
The $D(3)$-brane effective dynamics can be obtained from the sigma model 
action for the open string, in a conformal gauge, with $\delta X^{\perp} =0$.
For constant closed string backgrounds:
metric $g_{\mu\nu}= 
g\delta_{\mu\nu}$, two-form $b_{\mu\nu}= b\eps_{\mu\nu}$, dilaton and a
constant $U(1)$ field strength $F_{\mu\nu}$ at the open string boundary, the
action \cite{fradkin} for $D(3)$-brane becomes 
\be
S\ =\ - {1\over{4\pi\a'}}\ \left [\ \int_{\S}\ d\tau\ d\s \ \pr_a X^{\mu}
\pr^{a}X^{\nu} \ g_{\mu\nu}\  -\ {\bar\caf}_{\mu\nu}\ \int_{\pr\S}\
d\tau\ X^{\mu}\pr_{\tau}X^{\nu} \ \right ]\ . \label{action}
\ee
${\bar\caf}_{\mu\nu} = (2\pi\a') {\caf}_{\mu\nu}
 = {( 2\pi\a' F + b )}_{\mu\nu}$ denotes the $U(1)$ gauge
invariant field strength.
The inherent ingredients in 
${\bar\caf}_{\mu\nu}$ are the gauge field $A_{\mu}(X)$ and the global two-form
$b_{\mu\nu}$. The latter is closed but not an exact form. Since a local transformation cannot
gauge away the global field, the $U(1)$ gauge symmetry is spontaneously broken in the 
theory. As a result,
$D(3)$-brane world-volume dynamics is described by a noncommutative $U(1)$ symmetry
\cite{seiberg}.

\sp
The string coordinates $X^{\mu}$ satisfies the world-sheet equation of motion 
$\Box X^{\mu} =0$ 
in bulk, as well as the mixed boundary condition
on the boundary{\footnote{For an infinite strip world-sheet topology, the
boundary is described at $\s = 0$ and $\pi$. In fact, the boundary possesses stringy
features and only to leading order in $\a'$, the boundary describes points there.}}
\be
g_{\mu\nu}\ \pr_{\s}X^{\nu}\ -\ {\bar\caf}_{\mu\nu}\ \ \pr_{\tau}X^{\nu}\
=\ 0 \ ,\label{bc}
\ee
where ${\bar\caf}_{\mu\nu}={\bar\caf} \eps_{\mu\nu}$ is an invertible matrix.
The string mode expansion for its coordinates become
\bea
X^{\mu}(\tau,\s)&=& x^{\mu}\ +\ 2\a' \ p^{\mu}\tau 
\ +\ {1\over{\pi}}({\caf}^{-1})^{\mu\nu}\ p_{\nu}\s \nonumber\\
&&{}\qquad\quad +\ {\sqrt{2\a'}} \ 
\sum_{n\neq 0} {{e^{-in\tau}}\over{n}}\ \left ( 
\ i \ \a^{\mu}_n\ \cos n\sigma\ 
\ + {\bar\caf}^{\mu}_{\nu}\a^{\nu}_n\
\sin n\s\ \right ) .\label{modes}
\eea
The mode expansion for the string, contains two
independent perturbation parameters ($\a'$ and {$\caf$}). However on the 
$D(3)$-brane world-volume, the effective dynamics is described by a single effective
parameter $\a'_{\rm eff}$\cite{seiberg}.
We use the notations {${\caf}_{0i}= E_i$} for electric
components and {${\caf}_{ij}= \eps_{ijk} B_k$} ($i= 1,2,3$)
for magnetic ones in the theory. Explicitly, the
EM-field strength can be expressed
\be
{{\caf}}_{\mu\nu}\ =\ \left ( \matrix {  {0} & {E_1} & {E_2} & {E_3}\cr
                     {-E_1} & {0} & {B_3} & {-B_2}\cr
                     {-E_2} & {-B_3} & {0} & {B_1}\cr
                     {-E_3} & {B_2} & {-B_1} & {0} \cr }\right )
\ .\label{em-comp}
\ee
The Poincare dual ${}^{\star}{\caf}^{\mu\nu} = {1\over2} \eps^{\mu\nu\rho\sigma}
{\caf}_{\rho\sigma}$ can be expressed
\be
{{}^{\star}{{\caf}}}^{\mu\nu}\ =\ \left ( \matrix { {0} & {B_1} & {B_2} & {B_3}\cr
                     {-B_1} & {0} & {-E_3} & {E_2}\cr
                     {-B_2} & {E_3} & {0} & {-E_1}\cr
                     {-B_3} & {E_2} & {-E_1} & {0} \cr }\right )
\ .\label{dualem}
\ee
The Lorentz invariants on the $D(3)$-brane world-volume are given by 
\bea
&&{}^{\star}{\caf}_{\mu\nu} {\caf}^{\mu\nu} \ = - 4 \left ( E\cdot B\right )
\nonumber\\
{\rm and}\qquad &&{\caf}_{\mu\nu} {\caf}^{\mu\nu}\ = - 2 \left (E^2 - B^2\right ) \ .
\label{inv}
\eea
Since the direction of $E$-field is all along the string,
an electric stretch is parallel (or anti-parallel) to that of string. 
The stretch builds up an induced (electric) string tension and hence a
length scale, smaller in magnitude,{\footnote{Since the
noncommutative scale is a short-distance probe for testing string scale.}}
to that of the string scale $\a'$ in
the theory. As a consequence, the tension of the fundamental string receives a
correction from that of an $E$-field. A priori, the effective string tension can be
described by 
\be
{T}^2_{\rm eff}\ =\ {T}^2\ \pm\ {E}^2 \ .\label{vectorT}
\ee
The electric string alone gives rise to a new stringy description, coupled to the
conventional string and in a boundary description
it corresponds to NCOS.

\sp
On the other hand, the $B$-field induces a "stretch" along the transverse
direction to string and sets up a magnetic scale (independent of electric
one) in the theory. Needless to mention that the magnetic scale does not contribute
to string tension, since no work is done by a $B$-field alone. Nevertheless,
the $B$-field contributes substantially to energy along a transverse direction
to propagation. As a result, the magnitude of $T_{\rm eff}$ (\ref{vectorT}) is scaled by an 
overall constant 
factor due to the transverse growth.

\subsection{Bound on electric (magnetic) field}

\hs 
The fact that space-like noncommutative field theories are
perturbatively unitary at one loop level \cite{minwalla} and the time-like
ones are not \cite{gomis}, confirms different roles played by the $E-$ and 
$B$-fields in the theory.
We recall \cite{connes}-\cite{chu1} the effective noncommutative
$[ X^{\mu}, X^{\nu} ] = i \Theta^{\mu\nu}$ description on the world-volume of a
$D(3)$-brane. There, the momentum square can be given by
the inner product 
\be
P_{\rm nc}\cdot P_{\rm nc}\ =\ - p_{\lambda}\Theta^{\lambda\mu} \Theta_{\mu\rho}
p^{\rho} \ .\label{ncp}
\ee
The constant matrix ${\Theta}^{\mu\nu}$ can be written in block-diagonal form,
without loss of generality{\footnote{The block-diagonal form for the
non-commutative matrix parameter is obtained with Euclidean signatures and an
analytic continuation is performed to obtain ${\Theta}^{\mu\nu}$ in terms of
electric $\Theta_E$ and magnetic $\Theta_B$ parameters.}}
\be
\Theta^{\mu\nu}\ =\ \left ( \matrix { {0} & {\Theta_E} & {0} & {0} \cr 
{- \Theta_E} & {0} & {0} & {0} \cr {0} & {0} & {0} & {\Theta_B} \cr
{0} & {0} & {-\Theta_B} & {0} \cr }\right ) \ .\label{block}  
\ee
Simplification in terms of 
electric $\Theta_E$ and magnetic $\Theta_B$ parameters yield
\be 
P^2_{\rm nc}\ =\ \Theta_E \left ( p^2_0 - p^2_1 \right ) \ +\ 
\Theta_B \left ( p^2_2 + p^2_3 \right ) \ .\label{ncP}
\ee
For time-like non-commutativity ($i.e.\ \Theta_E\neq 0$), 
$P^2_{\rm nc}$ can be negative. The presence of unphysical states lead to
a description with non-unitary S-matrix.{\footnote{For space-like
non-commutativity ($i.e.\ \Theta_B\neq 0,\ \Theta_E = 0$), the theory
describes physical states, $P^2_{\rm nc} > 0$.}} However a 
bound on $E$-field is known to resolve the problem of unitarity in
the theory at the expense of the field theoretic dynamics. The precise bound on 
$E$-field may be obtained by confining to a physical domain for the brane
dynamics, which is described
by DBI Lagrangian density. For a $D(3)$-brane, it can be simplified to yield
\be
{\cal L}_{\rm DBI} \ =\ - T_{D}\ {\left [ g^4 \ +\ g^2\ (2\pi\a')^2 \
(B^2 - E^2)\ - \ (2\pi\a')^4 \ (E\cdot B)^2 \right ]}^{1/2} \ . \label{DBI}
\ee
Here $T_D= 1/[g_s (2\pi )^3 {\a'}^2]$ denotes the $D(3)$-brane 
tension, when $g_s$ corresponds to
the closed string coupling. In the weak coupling limit $g_s<<1$, the $D(3)$-brane
becomes heavy and signifies a smaller probe than the fundamental string.
It is important to note that the physical
domain in Minkowski space-time is represented by a positive and 
definite quantity
under the square root in eq.(\ref{DBI}). In the case, when both the Lorentz invariants
(\ref{inv}) are set to zero, the DBI Lagrangian density confirms no bound
on $E$-field. On the other hand, when one or both of the invariants
are non-zero, one obtains an upper bound on $E$-field. For instance
when $E\perp B$, the bound on $E$-field is less restrictive.
There, the relative strength between $E$- and $B$-fields decides
on the cut-off value. It is interesting to note that the bound on 
$E$-field takes an unique value for all cases. The critical value can
be obtained from eq.(\ref{DBI})
\be
E_c\ =\ {{g}\over{(2\pi\a')}} \ .
\label{bound}
\ee
It corresponds to that of string tension $T$ in the theory. There is no natural
bound on $B$-field, though under certain circumstances, $|B| < |E|$, $B$-field is 
bounded.
As $E$ increases, the length scale for the electric string also 
increases. At $E_c$, it becomes maximum and 
is equal to that of the fundamental string.
Since all the closed string modes decouple
in NCOS limit \cite{susskind, gopa1}, the residual open string becomes rigid, $i.e.\
T \ra \infty$. However, the effective string becomes tension-less $T_{\rm eff}\ra 0$.
In other words, an $E$-field drains off the string tension for a
(maximal) stable configuration. Then the effective string tension
(\ref{vectorT}) can precisely be described  by
\be
{{\a'}\over{\a'_{\rm eff}}}\ = \ \Big ( {{g^2}\ -\ |2\pi \a' E|^2}\Big )^{1\over{2}}
\ .\label{ratio}
\ee
In the limit $E\ra E_c$, the relative strength
between the length scales becomes ${{\a'}/{\a'_{\rm eff}}}\ra 0$, when
$\a'_{\rm eff}$ and the effective open string coupling are kept fixed.
Though the closed string coupling $g_s= ({\a'_{\rm eff}/{\a'}})^{1/2} \ra\infty$,
it remains in the bulk and decouples from the string boundary, $i.e.$ from
the $D(3)$-brane world-volume dynamics. 

\sp
In this context, it is useful to recall that a $D(3)$-brane forms a bound state 
\cite{witten,gukov} with
the fundamental string in presence of an $E$-field. The effective tension
for the bound state can be described by 
\be
T^2_{\rm eff}\ =\ T^2_D\ +\ {\tilde E}^2 \ .\label{boundT}
\ee
In the decoupling limit $T_{\rm eff}\ \ra 0$, 
the critical electric field can be identified with the
$D(3)$-brane tension, ${\tilde E}_c = i T_D$. It implies that the
$D(3)$-brane tension becomes imaginary in the gravity decoupling limit
${\tilde E}\ra {\tilde E}_c$. There, the massless closed string modes
are cancelled by the electric string and at the same time, the number of "rigid" strings
becomes large. They join end to end in the direction of $E$-field and describe a 
"long" NCOS in the theory. 

\subsection{Unitarity}

\hs
In this section, we comment on unitarity in a space-time non-commutative
gauge theoretic regime in a NCOS theory.
We obtain the Hamiltonian
in the theory by  taking into account the longitudinal and transverse momenta.
Some of the light string modes become
prominent in the theory and can be seen to modify the Hamiltonian.

\sp
Now consider parallel $E$ and $B$ fields in a open bosonic
string theory.
The total momentum ${\cal P}$ for the string, at an instant ($\tau=0, \s=0$),
in presence of an EM-field 
can be expressed in terms of the momenta,
along the direction of motion
$P_{\mu}^{\parallel}(\tau,\s)$ and along the transverse direction
$P_{\mu}^{\perp}(\tau,\s)$. It becomes
\be
{\cal P}_{\mu}\ =\ \left [ {\int}_{0}^{\pi} d\s P_{\mu}^{l}\right ]_{\tau =0} 
\ +\ \left [{\int}_{0}^{\infty}
d\tau P_{\mu}^{\perp}\right ]_{\s=0} \ .\label{totalP}
\ee
The longitudinal and transverse momenta can be obtained 
from the Lagrangian density in 
eq.(\ref{action}). They are
\bea
&&P_{\mu}^{l}\ =\ {1\over{2\pi\a'}} \left ( g_{\mu\nu} \pr_{\tau} X^{\nu}\
+\ {\bar\caf}_{\mu\nu} \pr_{\s} X^{\nu}\right )
\nonumber\\
{\rm and}\qquad &&P_{\mu}^{\perp}\ =\ - {1\over{2\pi\a'}}
\left (g_{\mu\nu} \pr_{\s} X^{\nu}\ +\ {\bar\caf}_{\mu\nu} \pr_{\tau} X^{\nu}
\right ) \ .\label{mom}
\eea
The total longitudinal momenta can be obtained by integrating over the infinitesimal
flow. The longitudinal momenta consist of a component ${\cal P}_0^l$ along the
propagation and ${\cal P}_1^l$ along the $E$-field. The remaining
two components (${\cal P}_2^l$, ${\cal P}_3^l$) are transverse to 
(${\cal P}_0^l$, ${\cal P}_1^l$).
Explicitly, the momenta become
\bea 
{\cal P}_{0}^{l}&=& - \left ( {g - g^{-1}{\bar E}^2} \right ) p^0 \ +
\ i (4 E)\ {\sqrt{2\a'}} 
\sum_{n\neq 0}\ {1\over{(2n+1)}} \a^1_{2n+1}
\ ,\nonumber\\
{\cal P}_{1}^{l}&=& \left ( {g - g^{-1}{\bar E}^2} \right ) p^1 \ -
\ i (4 E)\ {\sqrt{2\a'}} 
\sum_{n\neq 0}\ {1\over{(2n+1)}} \a^0_{2n+1}
\ ,\nonumber\\
{\cal P}_{2}^{l}&=& \left ( {g - g^{-1}{\bar B}^2} \right ) p^2 \ +
\ i (4 B)\ {\sqrt{2\a'}} 
\sum_{n\neq 0}\ {1\over{(2n+1)}} \a^3_{2n+1}
\ ,\nonumber\\
{\cal P}_{3}^{l}&=& \left ( {g - g^{-1}{\bar B}^2} \right ) p^3 \ -
\ i (4 B)\ {\sqrt{2\a'}} 
\sum_{n\neq 0}\ {1\over{(2n+1)}} \a^2_{2n+1}
\ .\label{mommodes}
\eea
The corrections to constant momentum modes are associated with the EM-field. They are
the constant and non-zero momentum corrections and as a whole describes a complex 
longitudinal momenta. Intuitively,
the longitudinal momenta describes a space-time 
non-commutative "field" theoretic regime in the string theory. The complex momenta (\ref{mommodes}) 
in the regime, describes negative norm states in addition to the physical states 
and lead to a non-unitary description.
Nevertheless, there are transverse flow of momenta in the theory 
due to the stringy description. It can be checked that
the transverse momenta ${\cal P}^{\perp}_{\mu}$,
cancel the imaginary non-zero modes in longitudinal momenta.
The total momentum ${\cal P} = {\cal P}^l + {\cal P}^{\perp}$ becomes
proportional to that of the constant momentum modes ({\ref{mommodes}). The proportionality
constant truns out to be the modified metric $g_{\mu\nu}\ra (g_{\mu\nu} - {\bar\caf}_{\mu\l}g^{\l\rho}
{\bar\caf}_{\rho\nu})$ in the NCOS theory.
Thus the EM-string theory describes the 
propagation of a non-gravitating "rigid" string.
Interestingly, the absence of closed string poles 
have been confirmed 
in the non-planar one-loop diagram in NCOS theory \cite{gopa1}. Though
the presence of graviton pole in the planar one-loop diagram is essential to
retain the unitarity in open string theory, a NCOS theory does not require so.

\sp
Now the Hamiltonian describing the propagation of string in presence of
EM-field can be given
\be
H\ =\ {\int}_{0}^{\pi}\ d\s\ \left (\ P^{\parallel}_{\mu}\
\pr_{\tau}X^{\mu} \ +\
P^{\perp}_{\mu}\ \pr_{\s} X^{\mu}\ - \ {\cal L}\ \right ) 
\ ,\label{Hamilton}
\ee
where the Lagrangian ${\cal L}$ can be obtained from the sigma model action
(\ref{action}). The Hamiltonian ($H =L^b_0$, zero-mode for Virasoro generators in
presence of a constant $b$-field) 
becomes
\be
L^b_0\ =\ \a' \ p^2 \ -\ \a' \
\left ({\bar\caf}g^{-1}{\bar\caf}\right )_{\mu\nu} p^{\mu}p^{\nu}
\ +\ N_{\rm osc}
\ , \label{exH}
\ee
where $N_{\rm osc}$ denotes the number operator and its eigen values are ($0,1,2, \dots$)
in the theory. 
The Hamiltonian receives a correction due to the light string modes in the theory.
Then the mass-shell condition  becomes
\be
\left (\ L^b_0\ -\ 1\ \right ) \left | {\rm phy}\right >\ =\ 0\ .\label{onshell}
\ee
The poles in the theory describe physical as well as unphysical states. However the
unphysical states are orthogonal to the physical ones. The orthogonality condition
can be implemented to get rid of the unphysical states from the NCOS theory. We postpone
the relevance of mass-shell condition to sections 3 and 4.

\section{New string modes: Small time-like dimension}

\subsection{$E\perp B$ }

\hs
Consider an EM-field, with $E = ( E_1,0,0 )$ and $B = ( 0, B_2, 0 )$,
compatible with perpendicular electric and magnetic field
configurations $E\cdot B =0$. Any other combination
for the field components satisfying $E\cdot B =0$, may be obtained from the
above case. For instance, if one switches on $E_3$ (or $B_3$) in addition to $E_1$ 
and $B_2$, the electric (magnetic) components can be rotated to yield
a single independent component. 

\sp
\noindent
The boundary conditions (\ref{bc}) for the perpendicular fields become
\bea
&& \pr_{\s} X^0\ -\ {{E}\over{E_c}}\ \pr_{\tau} X^1\ = 0 \ ,\nonumber\\
&& \pr_{\s} X^1\ +\ {{E}\over{E_c}}
\ \pr_{\tau} X^0\ +\ {{B}\over{E_c}}\ \pr_{\tau} X^3
\ = 0 \ ,\nonumber\\
&& \pr_{\s} X^2\ = 0 \nonumber\\
{\rm and}\quad 
&& \pr_{\s} X^3\ -\ {{B}\over{E_c}}\ \pr_{\tau} X^1\ = 0 \ .\label{perpbc}
\eea
\sp
Explicitly, the mode expansion (\ref{modes}) in the case becomes
\bea
&&X^0 =\ x^0 \ +\ 2\a'\ p^0 \tau\ +\ {1\over{\pi}} g E^{-1}\ p^1 \s \nonumber\\
&&\qquad\quad +\ 
{\sqrt{2\a'}}\ \sum_{n\neq 0}\ {{e^{-i n\tau}}\over{n}}\ \left (\ i\ \a^0_n\ 
\cos n\s\ +\ g {\bar E}^{-1}\ \a^1_n\ \sin n\s\ \right )\ ,\nonumber\\
&&X^1 =\ x^1 \ +\ 2\a'\ p^1 \tau\ -\ {1\over{\pi}} \left ( g E^{-1}\ p^0 \ +\
g B^{-1}\ p^3\right ) \s \nonumber\\
&&\qquad\quad
+\ {\sqrt{2\a'}}\ \sum_{n\neq 0}\ {{e^{-in\tau}}\over{n}}\ \left (\ i \a^1_n\ 
\cos n\s\ -\ \left ( g {\bar E}^{-1}\ \a^0_n\ +\ g {\bar B}^{-1}\ \a^3_n 
\right )\ \sin n\s\ \right )\ ,\nonumber\\
&&X^2 =\  x^2 \ +\ 2\a'\ p^2 \tau\ 
\ +\ i\ {\sqrt{2\a'}}\ \sum_{n\neq 0}\ {{e^{-in\tau}}\over{n}}\ \a^2_n\ 
\cos n\s\ \nonumber\\
{\rm and}\qquad&&X^3 = \ x^3 \ +\ 2\a'\ p^3 \tau\ +\ 
{1\over{\pi}} g B^{-1}\ p^1 \s \nonumber\\
&&\qquad\quad
+\ {\sqrt{2\a'}}\ \sum_{n\neq 0}\ {{e^{-in\tau}}\over{n}}\ \left (\ i \a^3_n\ 
\cos n\s\ +\ g {\bar B}^{-1}\ \a^1_n\ \sin n\s\ \right )\ .
\label{permodes}
\eea
The additional string modes in the spectrum are entirely due to the 
EM-field. They describe a transverse growth during the string propagation 
and may be interpreted as the light mass winding modes due to its association
with the T-dual description
in the effective NCOS theory. In other words, the
new string modes are unlike to that of a massless closed string windings in the theory.
The light string modes are reflected by the effective metric
in the NCOS theory, which may possess its root in the light mass closed string mode. 
In fact, the strange behaviour is inherited in the Hamiltonian (\ref{exH}).
In the effective description, the winding and momentum modes simultaneously contribute
to the Hamiltonian due to the noncommutative constraint, which in turn can be seen to
describe a $T$-duality in the NCOS theory.

\sp
In the case, a variation in $E$-field is independent from that in $B$-field
and vice-versa. However, the
relative strength between the fields
decides over the fate of the gravity decoupled theory. 
There are three
distinct cases: 
i).$|E| = |B|$, ii). $|E| < |B|$ and iii). $|E| > |B|$. 
For an equal field strength, eq.(\ref{DBI}) confirms no bound  and
hence a string decoupling limit (large $E$ and large $B$) can be 
reached. It corresponds to a light-like
noncommutative field theory \cite{aharony}.
For the second case ($|E|<|B|$), there is no bound on the fields,
which in turn leads to a space-like noncommutative field theory in the string
decoupling limit. Unlike the (ordinary) field theory, the noncommutative one
possesses winding modes and shares some of the properties of a
thermal field theory \cite{fischler}.
In this case, the $E$-field is not bounded by the string tension rather
it is bounded by the $B$-field strength. For a strong $B$-field
(large $B$ limit), $E$-field can also take a large value. The boundary conditions
(\ref{perpbc}) to leading order in $B$-field yields field theoretic modes
\be
X^1\ \ra\ x^1\qquad\qquad 
{\rm and}\qquad\quad X^3\ \ra\ x^3\ .\label{ft1}
\ee
Since some of the magnetic field components ($B_1=0$ and $B_3=0$) are zero,
the complete gravity decoupling limit for $E\perp B$ becomes $\a'\ \ra 0$ and
large $B$ (when $\a'$ is kept fixed). For $\a'\ \ra 0$, all the oscillator modes
drop out from the spectrum including the ones from $X^0$ and $X^2$
\be
X^0\ \ra \ x^0 \qquad\qquad
{\rm and}\qquad\quad X^2\ \ra\ x^2 \ .\label{qft2}
\ee

\sp
\noindent
In the third case
$|E|>|B|$, due to a bound on $E$-field (\ref{bound}), the string
decoupling limit is replaced by that of gravity 
and describes a time-like noncommutative theory.
Though a variation in $E$-field does not affect the $B$-field,
in the decoupling limit, the magnitude of magnetic scale
is bounded by that of electric, which in turn coincides with the intrinsic
scale in the theory. Then the gravity decoupling limit
is described by 
\be
g E^{-1}\ \ra \ 2\pi \a'\qquad\qquad {\rm and} \qquad\quad g B^{-1}\ \ra \
2\pi \a'_{\perp} \ ,\label{limit1}
\ee
where $\a'$ and $\a'_{\perp}$, respectively, denote the "rigid" scales for electric string
and transverse (magnetic) width in the theory. Interestingly, the "rigid"
electric (magnetic) string can be identified with the electric (magnetic) dipole there.
The EM-string modes in the limit can be obtained from eq.(\ref{modes}).
They are
\bea
&&X^0 =\ x^0\ +\ 2\a'p^0\tau \ +\ 2\a' p^1 \s
\ +\ {\sqrt{2\a'}} 
\sum_{n\neq 0}
{{e^{-in\tau}}\over{n}}
\left (i \a^{0}_n\ \cos n\s\ +\
\a^1_n \sin n\s \right )\ , \nonumber\\
&&X^1 =\ x^1\ +\ 2\a' p^1\tau\ - 
\ 2\a' p^0\s 
\ +\ {\sqrt{2\a'}}
\sum_{n\neq 0}
{{e^{-in\tau}}\over{n}}
\left (
i \a^1_n\ \cos n\s\ -\ \a^0_n\ \sin n\s\ \right )
\nonumber\\
&&\qquad\qquad\qquad\qquad\qquad -\ 2\a'_{\perp} p^3 \s
\ -\ {\sqrt{2\a'_{\perp}}} \sum_{n\neq 0}
{{e^{-in\tau}}\over{n}}
\ \a^3_n\ \sin n\s
\ ,\nonumber\\
&&X^2 =\ x^2\ +\ 2 \a' p^2\tau\ + 
\ i\ {\sqrt{2\a'}}
\sum_{n\neq 0}
{{e^{-in\tau}}\over{n}}
\a^2_n\ \cos n\s\ 
\nonumber\\
{\rm and}\qquad&&X^3 =\ x^3\ +\ 2\a' p^3\tau\ 
+\ i\ {\sqrt{2\a'}}
\sum_{n\neq 0}
{{e^{-in\tau}}\over{n}}
\a^3_n\ \cos n\s\ \nonumber\\
&&\qquad\qquad\qquad\qquad\qquad
+\ 2\a'_{\perp} p^1 \s\
+ \ {\sqrt{2\a'_{\perp}}}
\sum_{n\neq 0}
{{e^{-in\tau}}\over{n}}
\ \a^1_n \sin n\s
\ .\label{ncmodes}
\eea
The presence of winding modes 
(for the brane coordinates $X^0$, $X^1$ and $X^3$) in a gravity decoupled theory may be
surprising at first sight. However, on the world-sheet boundary the string modes 
(\ref{ncmodes}) satisfy $X^{\mu}|_{\s={\pi}} \ra X^{\mu}|_{\s=0} + 2 \pi \a' \eps^{\mu\nu} p_{\nu}$.
Since both ends of the open string lie on the $D(3)$-brane, $|\a' p|$ may be identified 
with the winding radius for the corresponding brane coordinate. The winding modes becomes
transparent in the effective description, where the noncommutative constraint is described
by the $T$-duality in the theory.
As a result, the NCOS can be seen to be
tied with the unconventional "closed" string. 
 
\sp
The length scales ${\sqrt{2\a'}}= l_s$ and ${\sqrt{2\a'_{\perp}}} =l_{\perp}$ in 
eq(\ref{ncmodes}) are along the direction of $E$- and $B$-fields 
($l_s <l_{\perp}$) and they signify string and magnetic
scales, respectively, in the theory. 
Different scales $l_s$ and $l_{\perp}$ confirm two independent
winding modes for $X^1$ with respective radii $R$ and $R_{\perp}$.
Identification of the EM-string winding modes in eq.(\ref{ncmodes})
under T-duality, interrelates the radii. 
Then the string modes become
\bea
&& X^0 \ \ra \ X^0\ +\ m \Big ( 2 \pi R\Big ) \ ,\nonumber\\
&& X^1 \ \ra \ X^1\ +\ n \Big ( 2 \pi R^d \Big )\ +
 \ n' \Big (2 \pi {{\a'}\over{\a'_{\perp}}} R^d_{\perp}\Big
)\ ,\nonumber\\
&& X^2\ \ra\ X^2 \nonumber\\
{\rm and}\qquad&& X^3 \ \ra \ X^3\ + \ w  \Big (2 \pi 
{{\a'}\over{\a'_{\perp}}}{R_{\perp}}\Big )\ ,
\label{ncoswind}
\eea
where ($m, n, n', w$) are arbitrary integers 
and denote the respective winding numbers for the EM-string coordinates.
The dual winding radii are 
$R^d= \a' /R$ and $R^d_{\perp} = \a' / R_{\perp}$.
Though the radius for $X^3$ is $R_3=\big ({\a'}/{\a'_{\perp}}\big ) R_{\perp}$,
its ratio with $R^d_3$ is independent of the string scale $R_{\perp}/R^d_{\perp}$. 
The effective 
radius for $X^1$ becomes
\be
R_1 \ =\ {{\a'}\over{R}} \ +\ \Big ({{{\a'}^2}\over{\a'_{\perp}}}\Big ){1\over{R_{\perp}}} 
\ .\label{wind3}
\ee
Since $E > B$ implies $\big ({{{\a'}^2}/{\a'_{\perp}}}\big ) << 1$, 
the radius for $X^1$ becomes
equal to the T-dual of $X^0$. 
For a large winding radius, the corresponding string coordinate describes a 
soliton string state. Its T-dual coordinate becomes small and
describes a particle state there.

\sp
The most revealing feature of the new string mode analysis is the fact that
the time-like (along with some of the space-like) coordinate becomes compact 
in the gravity decoupling limit.
There, eq.(14)
confirms the presence of tachyonic states in the $D(3)$-brane spectrum with Lorentzian
signature of space and time. In an equivalent picture, one may view the brane spectrum
as a physical one at the expense of imaginary $E$-field, which in turn implies a
real $E$-field in Euclidean time. 
As a consequence,
the notion of temperature (${\cal T}$) becomes prominent, in the
decoupling limit. In other words,
the notion of time breaks down in the limit, leading to imaginary
time, which in turn gives rise to the notion of temperature in the theory.
We postpone further discussion on thermal string to section 4.

\sp
On the other hand, since the $U(1)$ field strength is a constant, the gauge degrees of freedom are 
intertwined with that of the string $A_{\mu}(X) = - {1\over{2}} {\caf}_{\mu\nu} X^{\nu}$. 
The light modes in gauge field introduces the notion of "gauge-string" in the theory. The gauge
field/string windings
may be obtained from eq.(\ref{ncoswind})
\bea
&&A_0\ \ra\ A_0\ -\  \left (
{{\pi E}\over{\a'}}\right ) {1\over{R}}\ ,\nonumber\\
&&A_1\ \ra\ A_1\ -\  \left ( \pi E
\right ) R\ -\ \left ( \pi \a'_{\perp} B\right ){1\over{R_{\perp}}} \ ,\nonumber\\
&&A_2\ \ra\ A_2 \nonumber\\
{\rm and}\qquad&&A_3\ \ra\ A_3\ +\ \left (\pi B\right )\ R_{\perp} \ .
\label{gaugewind1}
\eea
The gauge field windings imply a quantization of its conjugate momenta, $i.e$ 
the $E$-field is quantized{\footnote{For example, the
quantization of $E$-field on a $D$-string is discussed, in a different context,
in our earlier work \cite{kar3}.
In the present case, though
the $E$-field is on a $D(3)$-brane world-volume, it possesses a single component.}}
and effectively results in quantization of the string
slope $\a'_{\rm eff}
= n\a'$, where $\a'$ is the minimal scale. Intuitively, the
tension-less effective string at $E_c$ is made from a large number of
"rigid" EM-strings joined end to end and are aligned along the 
$E$-field. 
Since the $E$-field is along $X^1$, the effective NCOS
is oriented along a loop of radius $R_1$. The T-duality constraint
(\ref{wind3}) to leading order implies $R_1= R^d$ and can be seen to be associated with
an effective "closed" string description along $X^0$.

\sp
In order to obtain a bound on the time-like radius in the EM-string theory,
consider the Hamiltonian (\ref{exH}) for $E\perp B$. It can be written as
\be
L^b_0\ =\ L_0 \ +\ \left ( \a'\ +\ \a'_{\perp}\right ) (p_1)^2\ +\ {\a'_{\perp}}
(p_3)^2\ ,\label{perp1}
\ee
where $L_0\ = \a' p^2 + N_{\rm osc}$ is the Hamiltonian in absence of light string modes
(analogous to a free Hamiltonian) in the theory.
The mass-shell condition (\ref{onshell}), can be simplified for a winding state
(with winding number $\pm 1$), to yield 
\bea 
&&R\ =\ {\sqrt{\a'}}\ ,\nonumber\\
{\rm and}\qquad&&R_{\perp}\ =\ {\sqrt{\a'_{\perp}}}
\ .\label{radiiperp}
\eea
It implies that the 
radius for the time-like coordinate $X^0$ is of order of (electric) string
length $l_s$ and that of $X^3$ is of order of (magnetic) transverse width $l_{\perp}$.
Since $l_s < l_{\perp}$, the radius for $X^0$ is smaller than that for
$X^3$. Furthermore, a small time-like dimension in the theory implies that $X^1$ is large
(\ref{wind3}). 
Thus, the time-like coordinate possesses a large number of closely spaced
tiny intervals ($2\pi R$), which can be approximated as a
continuous (constant) parameter $m(2 \pi R)\ \ra\ C$. As a result, the
theory possesses a time-like killing vector $\big ({\pr}/{\pr X^0}\big )$ and is invariant
under a transformation
\be
X^0\ \ra \ X^0 \ + \ C
\ .\label{isometry}
\ee
The origin of time translation symmetry is in the $E$-field and makes sense only at 
the decoupling limit. The manifestation of time translation can
also be understood from T-duality. A priori, it reduces 
the dimension of space-time from $(3+1)$ to $3$-dimensions. However a careful mode
analysis implies a reduction in degrees of freedom by one unit 
in the constant mode sector, which is 
due to the interchange of winding and momentum modes.
It is established that the non-zero momentum modes in the theory, introduce
an additional continuous degree of freedom \cite{atick},
which accounts for the 4th dimension. 
Eventually, the degrees of freedom in the gauge-string theory remains 
unchanged.

\sp
On the other hand, a close look on the center of mass (cm) for the gauge
string coordinates may be useful to
explain the origin of continuous degrees of freedom from the non-zero momentum 
modes in the theory rather than the constant modes.
The cm-coordinates can be obtained for the string modes
\bea
&&x^0_{cm}=\ x^0 \ +\ \pi\a' p^1\ +  
\ {2\over{\pi}} {\sqrt{2\a'}} \sum_{n} {1\over{(2n+1)^2}}
\ \a^1_{(2n+1)} \ , \nonumber\\
&&x^1_{cm}=\ x^1\ + 
\ \pi\a' p^0\ 
\ +\ {2\over{\pi}} {\sqrt{2\a'}}
\sum_{n} {1\over{(2n+1)^2}}
\ {\a^0_{(2n+1)}}\nonumber\\
&&\qquad\qquad\quad+\ \pi \a'_{\perp} p^3 
\ +\ {2\over{\pi}} {\sqrt{2\a'_{\perp}}}
\sum_{n} {1\over{(2n+1)^2}}
\ \a^3_{(2n+1)}
\ ,\nonumber\\
&&x^2_{cm}= \ x^2\ ,\nonumber\\
&& x^3_{cm} =\ x^3\ - 
\ \pi \a'_{\perp} p^1 \
+\ {2\over{\pi}}{\sqrt{2\a'_{\perp}}}
\sum_{n} {1\over{(2n+1)^2}} \ \a^1_{(2n+1)} \ .\label{cm}
\eea
Firstly, the expressions account for the non-locality in time ($x^0$) and space 
($x^1$, $x^3$) coordinates, since the cm-coordinates satisfy $[ x^{\mu}_{cm} , 
x^{\nu}_{cm} ]=0$. Secondly, the time coordinate mixes with the spatial component
of the momentum modes and vice-versa. Intutively, the mixing of time is 
in the spirit of light-cone coordinates in field theory. However the subtlety is
due to the space-time non-commutativity, which is tied with the non-zero modes and
hence stringy description comes into picture. 
In addition,
the presence of constant as well as non-zero momentum modes in the expression for 
cm-coordinates is a new phenomenon in comparison to that of a
conventional string.
In fact, the non-zero momentum modes are the rarely spaced, constant,
light ones from the oscillators in the theory. Since under T-duality,
winding and momentum modes are interchanged, the number of degrees of freedom 
remains unchanged in the theory. The origin of
the light states in the cm-coordinates is due to the broken string world-sheet
parity. Under such a global discrete symmetry, the string modes become $x^{\mu}\ra x^{\mu}$,
$p_{\mu}\ra p_{\mu}$ and $\a^{\mu}_n\ra (-1)^n\a^{\mu}_n$. 
 It can be checked that under the string world-sheet symmetry
all the oscillator modes in 
$x^{\mu}_{\rm cm}$ are projected out. However in presence of an
$E$-field, such a symmetry is not respected. 

\subsection{$E\parallel B$}

\hs
In this section we consider a parallel
field{\footnote{The case for anti-parallel electric and magnetic fields 
can be obtained from the parallel case discussed. In fact, when $E$- and
$B$-fields are at an angle other than $\pi/2$, the analysis is analogous 
to that in the parallel case.}} 
configuration $E =( E_1, 0, 0 )$ and $B =( B_1, 0, 0 )$.
Here a bound on $E$-field becomes a bound on $B$-field as well. 
Since the difference with the perpendicular field configuration is
due to the orientation of $B$-field, the description for the time-like string
coordinate $X^0$ remains identical in both cases. Also, the boundary conditions 
(\ref{bc}) for $g_{00}$ and $g_{33}$ components are the ones,
obtained for perpendicular fields
in eq.(\ref{perpbc}). The remaining two boundary conditions can be given
\bea
&&\pr_{\s} X^1\ +\ {{E}\over{E_c}}
\ \pr_{\tau} X^0\ 
\ = 0 \nonumber\\
{\rm and}\quad 
&&\pr_{\s} X^2\ -\ {{B}\over{E_c}} \pr_{\tau} X^3\ = 0\ .\label{parabc}
\eea
\sp
The space-like string coordinates can be obtained from eq.(\ref{modes}). They are
\bea
&&X^1\ = \ x^1 \ +\ 2\a'\ p^1 \tau\ -\ {1\over{\pi}} g E^{-1}\ p^0 
\s \nonumber\\
&&\qquad\qquad\qquad\qquad\qquad
+\ {\sqrt{2\a'}}\ \sum_{n\neq 0}\ {{e^{-in\tau}}\over{n}}\ \left (\ i \a^1_n\ 
\cos n\s\ -\  g {\bar E}^{-1}\ \a^0_n  
\ \sin n\s\ \right )\ ,\nonumber\\
&&X^2\ = \ x^2 \ +\ 2\a'\ p^2 \tau\ +\ {1\over{\pi}} g B^{-1}\ p^3\s\nonumber\\
&&\qquad\qquad\qquad\qquad\qquad 
+\ {\sqrt{2\a'}}\ \sum_{n\neq 0}\ {{e^{-in\tau}}\over{n}}\ \left ( i \a^2_n\ 
\cos n\s\ +\ g {\bar B}^{-1} \a^3_n \sin n\s \right )\ ,\nonumber\\
&&
X^3\ = \ x^3 \ +\ 2\a'\ p^3 \tau\ -\ {1\over{\pi}} g B^{-1}\ p^2 \s \nonumber\\
&&\qquad\qquad\qquad\qquad\quad
+\ {\sqrt{2\a'}}\ \sum_{n\neq 0}\ {{e^{-in\tau}}\over{n}}\ \left (\ i \a^3_n\ 
\cos n\s\ -\ g {\bar B}^{-1}\ \a^2_n\ \sin n\s\ \right )\ .
\label{paramodes2}
\eea
In the case, all
the string coordinates possess light string modes, unlike to
that in the perpendicular field configuration. The modes for $X^0$
and $X^3$ remain unchanged, from that of $E\perp B$. On the other hand, the light string
modes for $X^1$ in the perpendicular field case is shared by the  $X^1$ and $X^2$ for
$E\parallel B$.
In the gravity decoupling limit (\ref{limit1}), the space-like gauge-string
modes become
\bea
&&X^1 = \ x^1 \ +\ 2\a'\ p^1 \tau\ -\ 2\a'\ p^0 
\s\ +\ {\sqrt{2\a'}}\ \sum_{n\neq 0}\ {{e^{-in\tau}}\over{n}}\ \left (\ i \a^1_n\ 
\cos n\s\ +\  \a^0_n  
\ \sin n\s\ \right )\ ,\nonumber\\
&&X^2 = \ x^2 \ +\ 2\a'\ p^2 \tau\ 
\ +\ i\ {\sqrt{2\a'}}\ \sum_{n\neq 0}\ {{e^{-in\tau}}\over{n}}\ \a^2_n\ 
\cos n\s\ \nonumber\\
&&\qquad\qquad\qquad\qquad\qquad\qquad\qquad
+\ 2\a'_{\perp}\ p^3\s
\ +\ {\sqrt{2\a'_{\perp}}}\ \sum_{n\neq 0}\ {{e^{-in\tau}}\over{n}}\ \a^3_n\ 
\sin n\s \ ,
\nonumber\\
&&X^3 = \ x^3 \ +\ 2\a'\ p^3 \tau\ 
+\ i\ {\sqrt{2\a'}}\ \sum_{n\neq 0}\ {{e^{-in\tau}}\over{n}}\ \a^3_n\ 
\cos n\s\ \nonumber\\
&&\qquad\qquad\qquad\qquad\qquad\qquad\qquad-\ 2\a'_{\perp}\ p^2 \s
\ -\ {\sqrt{2\a'_{\perp}}} \sum_{n\neq 0} {{e^{-in\tau}}\over{n}}\ \a^2_n\ \sin n\s \ .
\label{paramodes}
\eea
Since the massless windings are decoupled in the limit, the remaining
light windings are due to the EM-field.
There, the radius for $X^3$ is completely independent from that in $X^1$,
(unlike to that in $E\perp B$) rather it is
T-dual to that in $X^2$. Two of the space-like light modes in the theory are 
\bea
&&X^1\ \ra\ X^1\ +\ n (2 \pi R^d) \nonumber\\
{\rm and}\qquad\quad
&&X^2\ \ra\ X^2\ \pm\ n' \Big ( 2 \pi {{\a'}\over{\a'_{\perp}}} R^d_{\perp}
\Big )\ .
\label{parawind}
\eea
The modes for $X^0$ and $X^3$ remain identical to that
in $E\perp B$, obtained in eq.(\ref{ncoswind}). 
The radii for the transverse 
coordinates $X^2$ and $X^3$ are dependent on the relative strength 
$\big (\a'_{\perp}/{\a'}\big )$. 
Their ratio is independent of field strengths and can be
given by $\big (R^d_{\perp}/R_{\perp}\big )$. 
It confirms a "small" dimension to $X^3$ (or $X^2$)
and a large dimension to $X^2$ (or $X^3$) in the transverse space.
In addition, there are windings in the gauge field and they can be given by
\bea
&&A_0\ \ra\ A_0\ -\  \left (
{{\pi E}\over{\a'}}\right ) {1\over{R}}\ ,\nonumber\\
&&A_1\ \ra\ A_1\ -\  \left ( \pi E
\right ) R\ ,\nonumber\\
&&A_2\ \ra\ A_2\ -\  \left ( \pi \a'_{\perp} B\right )
{1\over{R_{\perp}}}\nonumber\\
{\rm and}\qquad&&A_3\ \ra\ A_3\ -\ \left (\pi B\right )\ R_{\perp} \ .
\label{gaugewind}
\eea
The presence of light string in the gauge field account for the
nonlocality in gauge theory. As discussed for $E\perp B$, the 
small radius $R$ in $A_1$, implies a
quantization of $E$-field and hence an quantized effective string scale $\a'_{\rm eff}$.

\sp
Now the Hamiltonian (\ref{exH}) for the gauge-string becomes
\be
L^b_0\ =\ L_0\ +\ \a'\ (p_1)^2\ +\ \a'_{\perp}\ (p_{\perp})^2\ ,\label{para1}
\ee
where $p_{\perp}$ denotes transverse momenta to $E$-field and to the
direction of propagation. The mass-shell condition (\ref{onshell}) for a
winding state ( with winding number $\pm 1$) becomes
\be 
\left ( \a'\ -\a'_{\perp}\right ) (p_1)^2\ +\ \a'_{\perp} (p_0)^2\ =\ 1 \ .
\label{paraonshell}
\ee
For an equal field strength $|E|=|B|$, the mass-shell condition can be further
simplified and one obtains a bound on the light-like radius $R={\sqrt{\a'}}$. As expected,
the radius $R$ is precisely the one obtained in eq.(\ref{radiiperp}) for
$E\perp B$.
A priori for $|E|\neq |B|$, the mass-shell condition (\ref{paraonshell}) can be
seen to introduce a different bound on the radius $R$. Nevertheless, in the
decoupling limit $\a'_{\perp} \ \ra\ \a'$, the radius
$R$ turns out to be of the order of string length $l_s$. 
The gauge-string scale for time-like intervals
in the limit, describes a time translation symmetry 
(\ref{isometry}) in the theory. The symmetry is intrinsic to the
$E$-field alone, at its critical value, and is independent of the orientation of
$B$-field in the theory. The non-zero mass
space-time windings introduce discrete space-time structure 
in the theory. However in the limit of critical $E$-field,
the time-like intervals are small enough in comparison to the space-like intervals.
In addition, at $E_c$, it gives
rise to a large number of time-like windings. 
As a result, the time coordinate can be approximated
as continuous rather than discrete. However due to the translation symmetry, the notion
of time decouples from the theory without affecting the degrees of freedom.
In other words, the time coordinate does not retain its identity and mixes with
the space components of non-zero momentum modes.
At the expense of time, the notion of temperature becomes prominent
in the theory.

\subsection{Mixing of time with space components}

\hs
In this section, we describe the interplay of time with space components of the
non-zero momentum modes in the gravity decoupling limit. 
A non-trivial mixing of time with space occurs in a gauge-string description with an arbitrary
orientation of $B$-field with respect to the $E$-field.
However a mixing among the space coordinates (rotation) takes place for non-perpendicular
fields. Taking account for a mixing in the momentum modes, separately in constant and in 
oscillator sector, the time-like coordinate in eq.(\ref{ncmodes})
can be re-expressed
\be
X^0 =\ x^0\ +\ 2\a' p^{-}\tau_{-} \ +\ 
2 \a' p^{+} \tau_{+}
\ +\ i {\sqrt{2\a'}} 
\sum_{n\neq 0}
{{e^{-in\tau_{-}}}\over{n}}\ \a_n^{-}
\ +\ i {\sqrt{2\a'}} 
\sum_{n\neq 0}
{{e^{-in\tau_{+}}}\over{n}}\ \a_n^{+}\ , \label{mommix}
\ee
$${\rm where}\qquad\quad 
p^{\pm}\ =\ {1\over{2}} \left ( p^0\ \pm\  p^1\right )\;\ ,\qquad 
\a_n^{\pm}\ =\ {1\over{2}} \left (\a_n^0\ \pm\ \a_n^1\right )
\qquad {\rm and }\qquad\tau_{\pm}= \tau \pm \s \ .$$
Similarly, one of the transverse space-like coordinates in eq.(\ref{paramodes})
can be re-expressed
\be
X^2 =\ x^2\ +\ 2\a'p^{s-}\tau_{-} \ +\ 
2 \a' p^{s+} \tau_{+}
\ +\ i {\sqrt{2\a'}} 
\sum_{n\neq 0}
{{e^{-in\tau_{-}}}\over{n}}\ \a_n^{s-}
\ +\ i {\sqrt{2\a'}} 
\sum_{n\neq 0}
{{e^{-in\tau_{+}}}\over{n}}\ \a_n^{s+}\ , \label{mommix2}
\ee
$${\rm where}\qquad\quad
p^{s\pm}\ =\ {1\over{2}} \left ( p^2\ \pm\ \Big ( {{\a'_{\perp}}\over{\a'}}\Big ) 
p^3\right )\;\ ,\qquad \a_n^{s\pm}\ =\ {1\over{2}} \left ( \a_n^2 \pm \Big 
({{\a'_{\perp}}\over{\a'}}\Big )^{1/2} \a_n^3 \right )\ .$$
The left-right independence of gauge string coordinate allows one to express
\be
X(\tau,\s )\ = \ X_L(\tau_{-})\ +\ X_R(\tau_{+}) \ .\label{lr} 
\ee
The left and right moving sectors in the open string 
coordinate is a new observation and is essentially due to the new string modes in the
theory. The winding energy in the theory is utilized completely 
to mix the notion of time with that of space. 
It is the 
$E$-field, which in the first place, drains off the closed string from the 
gauge-string spectrum in such a way that the effective string becomes tension-less.
Interestingly, in the process of draining off, the open string becomes "rigid" 
and a large number
of them (as E-field increases) 
join end to end to form a loop along the direction of $E$-field. In the
effective description, the "rigid" string configuration describes a tension-less
"long" string.

\sp
On the other hand, the expressions (\ref{mommix})-(\ref{mommix2}),
show that a mixing of time and space components, takes away all the
string modes in the left and right moving sectors. Then, the
zero-mode for the Virasoro generators become
\bea
&&L^{-}_0\ = {\a'} (p^{-})^2 \ + \ N^{-}_{\rm osc} \ ,\nonumber\\
&&L^{+}_0\ = {\a'} (p^{+})^2 \ + \ N^{+}_{\rm osc} \ ,\nonumber\\
&&L^{s-}_0\ = {\a'} (p^{s-})^2\ + \ N^{s-}_{\rm osc}\nonumber\\
{\rm and}\qquad&&
L^{s+}_0\ = {\a'} (p^{s+})^2\ + \ N^{s+}_{\rm osc} \ ,
 \label{vira}
\eea
where $N^{\pm}_{\rm osc}$ and $N^{s\pm}_{\rm osc}$ 
denote the left and right number operators in the theory.
The mass-shell condition (\ref{onshell}) can be re-expressed as
\be
{( \ L^{-}_0\ +\ L^{+}_0\ +\ L^{s-}_0\ +\ L^{s-}_0\ -\ 1\ )}|{\rm phy}>\ =\ 0
\ .\label{onshell1}
\ee
For constant modes, the 
mass-shell condition can be simplified in terms of the
original (un-mixed) string coordinates. Then the winding state with mass $M$,
for a single winding, can be given
\be
{{\a'}\over{2}} M^2 \ =\ {1\over{\a'}} R^2 \ -\ 1
\ .\label{mass-shell}
\ee
In the gravity decoupling limit, the
on-shell condition yields
$R= {\sqrt{\a'}}$, which is the radius for the time-like coordinate. Subsequent
decoupling of tachyonic modes confirms an upper bound to the dimension
of the time-like coordinate $R < {\sqrt{\a'}}$ for the gauge-string.

\section{Effective thermal string description}

\subsection{Hagedorn temperature}

\hs
In the previous section, we noticed that the 
notion of temperature is genuine in presence of an $E$-field.
Then, the gauge-string can be re-interpreted as a thermal string in the theory.
In this section, we analyze the gravity decoupling limit and subsequently, the 
light gauge-string(s) decoupling, in an alternate formulation of
thermal string theory. Interpreting the decoupling as a phase transition, we
investigate for the Hagedorn phase in the theory.

\sp
In the gravity decoupling limit, the effective string
scale $\a'_{\rm eff}$ can be obtained from the eq.(\ref{ratio}). In the limit 
$(g - {\bar E})\ \ra 0^{+}$ ($i.e.\ E\ra E_c$ from below), the string scale 
$\a'$ can be replaced by a large scale
$\a'_{\rm eff}$ and $\a'_{\perp} \ \ra (\a'_{\perp})_{\rm eff}$ in the theory.
The mode expansion for the gauge-string for $E\perp B$
(\ref{permodes}) can be 
re-expressed for a thermal
string. It becomes
\bea
&&X^0\ =\ x^0\ +\ {{(4\pi\a'_{\rm eff} )m'}\over{\beta}}\tau \ +\ {{m \beta}\over{\pi}} \s
\ +\ {\sqrt{2\a'_{\rm eff}}} 
\sum_{n\neq 0}
{{e^{-in\tau}}\over{n}}
\left (i \a^{0}_n\ \cos n\s\ +\
\a^1_n \sin n\s \right ) \ ,\nonumber\\
&&X^1\ =\ x^1\ +\ {{(4\pi\a'_{\rm eff}) n'}\over{\beta_d}}\tau\ - 
\ {{n\beta_d}\over{\pi}}\s \ 
-\ {{q \gamma_d}\over{\pi}}\s 
\ +\ {\sqrt{2(\a'_{\perp})_{\rm eff}}} \sum_{n\neq 0} {{e^{-in\tau}}\over{n}} \a^3_n \ \sin n\s
\nonumber\\
&&\qquad\qquad\qquad\qquad\qquad\qquad\qquad -\ {\sqrt{2\a'_{\rm eff}}}
\sum_{n\neq 0}
{{e^{-in\tau}}\over{n}}
\left (
i \a^1_n\ \cos n\s\ +\ \a^0_n\ \sin n\s \right )\ ,
\nonumber\\
&&X^2\ =\ x^2\ +\ {{(4\pi\a'_{\rm eff}) q'}\over{\gamma_d}} \tau \ +\ i{\sqrt{2\a'_{\rm eff}}}
{\sum_{n\neq 0}} 
{{e^{-in\tau}}\over{n}}
\a^2_n\ \cos n\s\ 
\nonumber\\
{\rm and}&&X^3\ =\ x^3\ +\ {{( 4\pi\a'_{\rm eff}) w'}\over{\gamma}}\tau + 
{{w \gamma}\over{\pi}} \s\ 
+ \ {\sqrt{2 (\a'_{\perp})_{\rm eff}}}
\sum_{n\neq 0}
{{e^{-in\tau}}\over{n}}
\ \a^1_n\  \sin n\s \nonumber\\
&&\qquad\qquad\qquad\qquad\qquad\qquad\qquad+ \ i\ {\sqrt{2\a'_{\rm eff}}}
\sum_{n\neq 0}
{{e^{-in\tau}}\over{n}}
\ \a^3_n\ \cos n\s\
\ ,\label{thermalperp}
\eea
where the thermal parameter $\beta =2 \pi R$ and the associated
parameter $\gamma = 2 \pi \big (\a'_{\perp}/{\a'}\big )_{\rm eff}
R_{\perp}$. Under S-duality, the parameters exchange their role
$\beta\leftrightarrow\gamma$ in the theory.
The thermal duality relation can be expressed
\be
\beta_d\ = \ {{4 \pi^2\a'_{\rm eff}}\over{\beta}}\ .\label{thermal}
\ee
Similarly, one can also obtain a duality relation involving $\gamma$-parameter in
the theory. In the closed string decoupling limit, $\beta$ is small and its dual
$\beta_d$ becomes large and vice-versa. It is in agreement with the observation 
that the winding radius for $X^1$ is large.

\sp
When the orientation of a $B$-field is changed by $\pi /2$, $i.e.$ 
$E\parallel B$, some of the space-like thermal string coordinates
(\ref{paramodes}) 
are different than that of $E\perp B$. They can be re-expressed in terms of thermal
string modes 
\bea
&&X^1 = x^1\ +\ {{(4\pi\a'_{\rm eff}) n'}\over{\beta_d}}\tau\ - 
\ {{n\beta_d}\over{\pi}}\s \
-\ {\sqrt{2\a'_{\rm eff}}}
\sum_{n\neq 0}
{{e^{-in\tau}}\over{n}}
\left (
i \a^1_n\ \cos n\s\ +\ \a^0_n\ \sin n\s \right )
\nonumber\\
{\rm and}\;\;&&X^2 = x^2\ +\ {{(4\pi\a'_{\rm eff}) q'}\over{\gamma_d}} \tau \
+\ {{q\gamma_d}\over{\pi}} \s\ +\ i{\sqrt{2\a'_{\rm eff}}}
{\sum_{n\neq 0}} 
{{e^{-in\tau}}\over{n}}
\a^2_n\ \cos n\s\ \nonumber\\
&&\qquad\qquad \qquad\qquad\qquad\qquad\qquad\qquad\qquad
+\ {\sqrt{2 (\a'_{\perp})_{\rm eff}}}
{\sum_{n\neq 0}} 
{{e^{-in\tau}}\over{n}}
\ \a^3_n\ \sin n\s
\ .\label{thermalpara}
\eea
The remaining mode expansion for $X^0$ and $X^3$, remains identical to that
of $E\perp B$ (\ref{thermalperp}).
The smallness of winding radius $R$ in the gravity decoupling limit gives rise
to a high temperature phase in the theory.

\sp
In either case ($E\perp B$ or $E\parallel B$), there are four light string modes
corresponding to space and time coordinates in the theory.
The mode expansion for thermal string confirms that the contribution to winding energy 
from $X^0$ (and $X^3$ for $E\perp B$) is  
equal to that for the momentum mode in $X^1$ (and $X^2$ for
$E\perp B$) due to the thermal duality symmetry (\ref{thermal}). 
Taking into account the winding
energy, 
the zero mode part of the Virasoro generator for open thermal string becomes
\bea
L^b_0&=& - {{\a'_{\rm eff}}\over{2}} M^2 \ +\ {\cal N}_{\rm osc} \
+ {{\a'_{\rm eff}}\over{2}} \left (
{{2\pi m'}\over{\beta}} \ +\ {{m\beta}\over{2\pi\a'_{\rm eff}}}\right )^2 \ 
+\ {{\a'_{\rm eff}}\over{2}} 
\left ({{2\pi n'}\over{\beta_d}}\ +\ {{n\beta_d}\over{2\pi\a'_{\rm eff}}}\right )^2
\nonumber\\
&&\qquad\qquad\qquad\;\; +\ {{\a'_{\rm eff}}\over{2}} 
\left ({{2\pi q'}\over{\gamma}}\ +\ {{q\gamma_d}\over{2\pi\a'_{\rm eff}}}
\right )^2\ +\ {{\a'_{\rm eff}}\over{2}}
\left ({{2\pi w'}\over{\gamma_d}}\ -\ {{w\gamma_d}\over{2\pi\a'_{\rm eff}}}
\right )^2 \ ,\label{vira2}
\eea
where ${\cal N}_{\rm osc}$ denotes the number operator. 
For a pure winding state with a single winding , the mass-shell (\ref{onshell})
condition yields
\be
{{\a'_{\rm eff}}\over{2}} M^2 \ =\\ {1\over{\a'_{\rm eff}}}\ \left ({{\beta}\over{2\pi}}
\right )^2 \ -\ 1
\ .\label{ms}
\ee
In the gravity decoupling limit, 
the free-energy vanishes $M=0$ and one obtains
a critical (Hagedorn) temperature 
\be
{\cal T}_H^{\rm eff}\ =\ {1\over{2\pi{\sqrt{\a'_{\rm eff}}}}} \ .\label{critT}
\ee
The effective Hagedorn temperature depends on the string scale and hence on a
phase in the theory. However a genuine Hagedorn temperature ${\cal T}_H$
is uniquely determined by the scale, where a transition leads to the 
Hagedorn phase. Figure 1, in section 4.3, shows a schematic demonstration of
the Hagedorn transition in presence of an $E$-field.
The effective string tension obtained from eq.(\ref{ratio}) can
be appropriately expressed in terms of temperature using eq.(\ref{critT}). 
Than the string tension becomes
\be
T_{\rm eff}\ =\ T \left ({{\cal T}_H^2\over{{\cal T}^2}}\ - 1 \right )^{1/2}
\ .\label{tension2}
\ee
Thus, the gravity decoupling limit, in thermal string descrption, is described by
${\cal T}\ra \ {\cal T}_H$.

\subsection{Critical regime: gauge-string(s) decoupling}

\hs
In this section, we shall notice that the Hagedorn limit ${\cal T}\ra {\cal T}_H$
does not correspond to a single sharp transition, rather it is associated with a
series of transitions in a critical regime. In fact, the phase transition begins 
in the gravity decoupling limit, passes through a large number of string phases
and leads to the Hagedorn phase. Let us qualify the statement by recalling the
Hagedorn transition in a fundamental string theory \cite{atick}. The mass-less
closed string backgrounds are the metric, an arbitrary two-form and a dilaton.
At this point, we recall the $U(1)$ gauge invariant combination 
${\bar\caf}= (b + {\bar F})$ for an arbitrary
$b$-field in the theory. The $b$-field and the gauge field belong to the mass-less
spectrum of closed and open string respectively. 
Taking account for the mass-less scalar (in a
higher order in $\a'$) and/or a mass-less pseudo-scalar (dual of two-form) the free 
energy eq.(\ref{ms}) can be seen to be lowered in the decoupling limit. In
other words, the transition takes place at a critical temperature below the 
Hagedorn temperature. A first order phase transition has been shown to take
place at a temperature, just below ${\cal T}_H$.

\sp
In the present context, the two-form $b$-field is a global mode of the
theory. It decouples from the bulk dynamics of the closed string and couples to
the open string boundary dynamics. Though an arbitrary $b$-field in the theory
has its origin in the massless closed string spectrum, a constant $b$-field
does not belong to the massless closed string spectrum any more, rather it is associated
with the $U(1)$ gauge interaction at the string boundary. The global modes of
$b$-field, modifies the $U(1)$ gauge invariance to a non-local one in the
theory. The spontaneous breaking of the local symmetry gives mass-term to the 
gauge field and the $b$-field is absorbed in an enhanced $U(1)$ symmetry
$i.e.$ a non-commutative $U(1)$ symmetry \cite{seiberg}. The non-local
extension is due to the non-linearity in EM-field and gives rise to gauge-string
description in the critical regime. 
Taking into account the non-zero mass gauge/string 
windings in eqs(\ref{gaugewind1}) and (\ref{gaugewind}), 
the effective critical (Hagedorn) temperature (\ref{ms}) is lowered than the
intrinsic Hagedorn temperature. As a result, the free
energy is lowered and the transition can be
seen to be associated with a discontinuity in free energy and describes a
first order transition. All the
closed string windings are favoured for decoupling
due to their vanishing mass during the transition.

\sp
Subsequently, as the temperature is increased in the regime (${\cal T}\ra {\cal T}_H$), 
at a certain
critical temperature $T_H^{\rm eff}$, it overcomes the threshhold and 
a fraction of non-zero mass string/gauge windings are decoupled from the theory.
Since the effective string scale $\a'_{\rm eff}= n\a'$ is quantized in the
unit of "rigid" scale $\a'$, 
the string scale in the effective description is
lowered $\big (\a'_{\rm eff}\big )^{\rm new} = (n-m)\a' < \a'_{\rm eff}$.
It implies that the critical (Hagedorn) temperature (\ref{critT}) in the theory increases
in each step, with the decoupling of light string from the gauge field in the theory.  
In other words, the critical
temperature ${\cal T}_H^{\rm eff}$ is self-tuned along with the decoupling of
gauge-strings at successive stages and the genuine Hagedorn transition recedes.

\sp
In order to analyze the nature of phase transition, we consider the partition function for
the thermal string in the critical regime. The Hamiltonian in the regime can be
obtained from eq.(\ref{vira2}) and the partition function becomes
\be
{\cal Z}\ =\ {\rm Tr}\ \exp \Big ({-\beta L_0^b}\Big ) \ .\label{partition1}
\ee
In the regime, the interaction Hamiltonian plays a leading role due
to significant contribution from the light string modes.
Explicitly, the expression (in the limit ${\cal T}\ra {\cal T}_H$ from below) becomes
\be
{\cal Z}\ =\ {\cal Z}_0\ {\rm Tr}\ 
\exp \left ({-\beta_H {L^w_0}}\right ) \ ,\label{partition2}
\ee
where ${\cal Z}_0$ denotes the partition function in absence of EM-field.
$L^w_0$ denotes the interaction Hamiltonian, which can be obtained from eq.(\ref{vira2}).
In the regime, the string energy is significantly large and $\beta_H$ is "fixed".
The expression (\ref{partition2}) describing the sum over gauge-string states,
converges and the total free energy ${\cal E} = -{\cal T} \ \ln {\cal Z}$ is analytic 
during the transitions. However, the specific heat ${\cal E}/({\cal T} - {\cal T}_H)$
shows nonanalytic behaviour in the critical regime and confirms a second order
transition. 
A series of discrete transitions of similar kind, have been shown to be 
second order due to a large number of
closely spaced first order transitions \cite{mukund,barbon2}.
Interestingly, in these papers, the 
second order phase transitions have been discussed, with an appropriate thermal ensemble.
There, the decoupling of strings at the transition, have been shown from the
bound state of $D(p)$-branes with F-strings in type II string theories. 

\sp
The critical regime (${\cal T}\ \ra {\cal T}_H$), is descibed by the phase area $ABC$ under
the curve depicted in figure 2, in section 4.3.
Various phases there, correspond to a thermal string, which 
comprises NCOS phase at its bottom (along $BA$). The mixing of time with space components
from the momentum modes begins in NCOS phase. The mixing is reflected due to the presence
of small mass (string $\leftrightarrow$ 
non-linear gauge) excitations in the phase. With an
increase in temperature, $i.e.$ along $B\ \ra C$ in the critical regime, the new light strings
begin to decouple from the theory. The fact that time dimension is small and 
the space-dimensions are large, imply a large winding number for time than the space 
coordinates. Thus the decoupled gauge-strings, during the second order
phase transition, comprises a significant number of time-like windings than the space-like
ones. In the case, a  mixed string state may be approximated by that of 
time-like ones. Intuitively, the decoupling of time-like string modes, during a series of 
transitions, partly drains off the non-linearity in the gauge field.
In this context, let us consider the $D(3)$-brane dynamics (\ref{DBI}) and recall some 
of the relevant mappings, established in the literatures, to gain some insight on the
high temperature phase in the theory. 

\sp
In the effective description, the DBI Lagrangian density (\ref{DBI}) can be expressed
as that for a noncommutative $U(1)$ to a leading order in
$\a'$. In the limit $\a'\ra 0$, the $D(3)$-brane dynamics reduces to $U(1)$ gauge
theory \cite{seiberg} with $\star$-product in its definition.
Schematically, ${\cal L}_{\rm DBI}\ \ra\
{\cal L}^{nc}_{\rm DBI}$ the noncommutative DBI Lagrangian density, which can be
expressed as
\be
{\cal L}^{nc}_{\rm DBI}\ =\ {1\over{8\pi G_s}}\ {\sqrt{\det G}}\ \ G^{\mu\l} 
G^{\nu\rho}\ Tr
F^{nc}_{\mu\nu} \star F^{nc}_{\l\rho} \ +\ {\rm total\; derivatives} \ . \label{ncMax}
\ee
Taking account for the temporal noncommutativity, the nonlinear relation 
between the field strength for a
noncommutative $U(1)$ field and that for (ordinary) gauge field
can be generalized from ref.\cite{seiberg}. To first order in
parameter $\Theta$ ( = $\Theta_E$ and $\Theta_B$), the noncommutative field strength
becomes
\be
F^{nc}_{\mu\nu}\ =\ F_{\mu\nu}\ +\ {\Theta}^{\l\rho}\ F_{\mu\l} F_{\nu\rho}\ 
-\ \Theta^{\l\rho}\ A_{\l}\ \pr_{\rho} 
F_{\mu\nu}\ .\label{relation2}
\ee
Now the noncommutative DBI Lagrangian ${\cal L}^{nc}_{\rm DBI}$ can be expressed
in terms of noncommutative $U(1)$ Lagrangian density with stringy corrections.
\be
{\cal L}^{nc}_{\rm DBI} =\ - {1\over{4}} \Big ( F^{nc}\Big )^2
\ +\ {\cal O}(\a' )\ ,\label{DBI5}
\ee
\bea
{\rm where}\;\; 
- {1\over{4}} \Big ( F^{nc}\Big )^2
&=& 
{1\over{2}} \Big ( E^2\ -\ B^2\Big ) \Big ( 1\ +\ \Theta_E\cdot E\ +\
\Theta_B\cdot B \Big ) \nonumber\\
&&\qquad\qquad -\ \Big ( E\cdot B\Big ) 
\Big ( \Theta_E\cdot B\ +\ \Theta_B
\cdot E \Big ) \ .\label{DBI6}
\eea
Since gravity is decoupled from the theory, the
noncommutativity essentially describes the stringy scale in the theory.
The nonlinearity in $U(1)$ theory
can be obtained from eq.(\ref{DBI6}). Interestingly, the trace of energy-momentum
tensor $T^{\mu}_{\mu}$ calculated in the gauge theory can be seen to be equal 
to the required nonlinearity in the dynamics. 
To ${\cal O}(\Theta )$, the the trace becomes
\be
T^{\mu}_{\mu}\ =\ \Big ( E^2\ -\ B^2\Big ) g 
\ -\ \Big ( E\cdot B\Big ) \Big ( {{\a'}\over{\a'_{\perp}}}\ +\ 
{{\a'_{\perp}}\over{\a'}} \Big )_{\rm eff} \ .\label{trace}
\ee
In the effective description, the EM-string corrections ${\cal O}(\a')$ cancel the
gauge nonlinearity and describes a unitary theory. 
For $E\perp B$, the nonlinearity in the $U(1)$ gauge theory can be
simplified to obtain $g |E|^2$
and for the parallel field configuration, 
it becomes $- 2 g |E|^2$ (or $-2 g |B|^2$). It implies that the non-linearity in
the $U(1)$ gauge theory is entangled with the EM-field. Figure 1 in section 4.3, 
shows that the gauge non-linearity is maximum at $A$, $i.e.$ 
in the beginning of the critical regime. It becomes feeble with the decoupling of gauge-strings
and ultimately takes a minimal value at ${\cal T}_H$. 
A careful analysis at this point suggests that the entanglement
possesses its grain in the global modes of $b$-field. At a minimal nonlinearity,
the $b$-field forms a condensate $\big < T_{\mu}^{\mu}\big >$, which is schematically
shown along the curve $CD$ in figure 2. 
As a consequence, the Hagedorn phase becomes free from the
gauge non-linearity and describes  an ordinary gauge theory 
$F^{nc} \ra F$. In the effective
string description, the noncommutative $U(1)$ symmetry is spontaneously broken to
the $U(1)$ gauge symmetry, by giving a vacuum expectation value to the $b$-field at
${\cal T}_H$.
The phenomenon is analogous to that of deconfinement in QCD. 
In presence of a large number ($N$) of coincident branes, the gauge group becomes
$U(N)$ and the theory describes a noncommutative $U(N)$ symmetry. There, the Hagedorn phase 
becomes subtle due to the presence of additional (non-abelian gauge) symmetry. Presumably,
a spontaneous breaking of noncommutative symmetry, followed by that of 
non-abelian symmetry \cite{polyakov} describes the deconfinement in QCD.

\subsection{Hagedorn phase: gauge-string condensate and tachyon decoupling}

\hs
In this section, we re-investigate the critical limit ${\cal T}\ra {\cal T}_H$
carefully to explain some of the important issues at the phase transition.
To be specific, the limit can be viewed as the one from below (left)
$i.e.$ $\big ( {\cal T}\ -
\ {\cal T}_H\big )\ra 0^{-}$ as well as the one from above (right)
$i.e.$ $\big ( {\cal T}\ - {\cal T}_H\big )
\ra 0^{+}$. In the (left) limit, $\a' \ \ra\ \a'_{\rm eff}$, $i.e.$ the electric
string description can be replaced by that of an effective thermal string. 
On the other hand, in the (right) limit the  string scale becomes $\a'\ \ra\
- \a'_{\rm eff}$. Then, the
thermal string mode expansion can be obtained from eq.(\ref{thermalperp}). It becomes
\be
X^0\ =\ x^0\ +\ {{(4\pi\a'_{\rm eff}) m'}\over{\beta}}\tau \ +\ {{m \beta}\over{\pi}} \s
\ -\ {\sqrt{2\a'_{\rm eff}}} 
\sum_{n\neq 0}
{{e^{-in\tau}}\over{n}}
\left ( 
 \a^{0}_n\ \cos n\s\ +\ i
\a^1_n \sin n\s \right ) \ .\label{tabove}
\ee
The constant momentum and winding modes, in the regime, remain unchanged from that
of left limit. However the oscillator sector has undergone an exchange
between the imaginary and real components. Then an effective
string state
with mass $M_s$ satisfies 
\be
- M_s^2\  =\ {{N}\over{\a'_{\rm eff}}} \ . \label{mst}
\ee
It implies that the effective gauge-string 
have experienced a critical transition, $i.e.\ \a'_{\rm eff} \leftrightarrow - \a'_{\rm eff}$,
at ${\cal T}_H$. An extrapolation for $\a'_{\rm eff}$ around ${\cal T}_H$ $i.e.$
from below to above and vice-versa, assures $\a'_{\rm eff} \ra 0$ at ${\cal T}_H$. Since
$\a'_{\rm eff}= (n-m) \a'$, such a phase
is associated with the decoupling of a large number of gauge-strings (large $m$), 
just below ${\cal T}_H$. It leads to the condensation 
$\big <T_{\mu}^{\mu}\big >$ of gauge-string in the effective description at ${\cal T}_H$. 
Thus, the gauge-string becomes heavy ${\cal T}_{\rm eff}
\ra {\rm large}$ at ${\cal T}_H$ and decouples from the phase. It is
followed by the decoupling of tachyonic particles/electric dipoles at ${\cal T}_H$ and 
leads to the Hagedorn phase. The total energy at ${\cal T}_H$ can be given by
\be
{\cal E}\ =\ \left ( T_{\rm eff} + {{N}\over{\a'_{\rm eff}}} \right ) l_s \ .\label{energy1}
\ee
For minimum energy ${d{\cal E}}/{dl_s}=0$. Then, a stable configuration is described by
$T_{\rm eff}=M_s^2$. 
Thus an increase in energy at ${\cal T}_H$, excites the minimal scale to a
large string scale, which becomes unstable and ultimately decays 
to a stable phase consisting of
non-interacting gauge particles. 

\sp
Now the partition function, in the limit of ${\cal T} \ra {\cal T}_H$ from above,
can be obtained from eq.(\ref{partition2}) by accounting the correct expression for
the Hamiltonian there. It can be expressed as
\be
{\cal Z}\ =\ {\cal Z}_0\ {\rm Tr}\ 
\exp \left ({\beta_H {L^w_0}}\right ) \ .\label{partition3}
\ee
The non-convergence of the partition function implies its break down. Then,
the free energy ${\cal E}= - {\cal T} \ln {\cal Z}$ at ${\cal T}_H$ shows a logarithmic divergence
and the Hagedorn phase is described by a high density of states at 
${\cal T}_H$. 
A schematic details on the Hagedorn transition(s), leading to the Hagedorn phase is
described in figure 2.

\sp
At this point, we recall the time translation symmetry in the critical regime
to analyze the exchange of notion between time and temperature.
By now, it is certain that the notion of time breaks down in the critical
regime. However at the expense of time, the notion of temperature becomes 
inevitable in the regime. Thus, the ($3+1$)-dimensional
space-time is described by a non-simplectic manifold $R^3\times S^1$ in the critical regime. 
Since a tree level world-sheet can not be described by $R^3\times S^1$ topology, it
confirms that the free energy should begin at one-loop order in the regime. However at 
${\cal T}_H$, the free energy can be estimated from eq.(\ref{ms}). 
Computation of free energy for the thermal string \cite{atick}, shows $1/g^2$ dependence at
${\cal T}_H$, where $g$ denotes the genus of the
world-sheet. It provides an evidence for the tree level world-sheet contribution
to the free energy at
${\cal T}_H$. Since the tree-level world-sheet is simply connected, the $R^3\times S^1$ 
does not remain valid in the Hagedorn phase. The time translational symmetry
is lost there, leading to a break down of temperature. The lost notion of time, in the
critical regime, is regained in the Hagedorn phase.

\begin{figure}[ht]
\begin{center}

\vspace*{-.5in}
\relax\noindent\hskip -8.4in\relax{\includegraphics{phase1.ps}}

\end{center}

\vspace*{4.5in}
\noindent {\bf Figure 1: Notion of temperature in presence of an EM-field.}

\vspace*{.1in}
\noindent
{ A time-like noncommutative string description can be equivalently 
described by that of thermal string. Various phases are obtained after decoupling all
the gravitons, gauge-strings, their condensate and tachyons.}

\end{figure}

\clearpage

\begin{figure}[ht]
\begin{center}

\vspace*{-.5in}
\relax\noindent\hskip -8.4in\relax{\includegraphics{phase.ps}}
\end{center}

\vspace*{4.8in}
\noindent {\bf Figure 2: Critical regime leading to Hagedorn phase.}

\sp
\noindent
{Phases involved in an effective thermal string description is shown by
considering the variation of string length with temperature. The area ABC
describes the critical regime for the thermal string. A first order phase transition
is shown along BA, where gravity is completely decoupled to yield a NCOS phase. As an
immediate feature, a series 
of second order phase transitions, at respective ${\cal T}_H^{\rm eff}$,
involving gauge-strings is shown from BA to C. In the limit of a critical temperature,
the new phase is described by a relatively small string scale and hence the decoupled
gauge-strings are "long" strings in the theory. A further increase in energy at ${\cal T}_H$, 
increases the gauge-string scale and describes the "long" string phase (CF). 
Such a "long" gauge-string phase is unstable and
decays to (ordinary) gauge particles (EF) by releasing excess energy in the guise of 
"long" strings (CD + DE). The gauge-strings condensate (CD), before they 
become tachyonic (DE). The flip between the notions of time and temperature may be
summarized in a table:

\vspace*{.2in}

{\large
$$\matrix {{\underline{\rm\bf Phase:}} & {} & {\underline{\rm\bf Time}} 
           & {\leftrightarrow} & 
           {\underline{\rm\bf Temperature}}\cr
           {} & {} & {} & {} & {} \cr
           {\rm Below\ AB:} & {} & {\rm valid}  & {} & {\rm valid}\cr
           {} & {} & {} & {} & {} \cr
           {\rm Area\ ABC:} & {} & {\rm breaks} & {} & {\rm valid}\cr
           {} & {} & {} & {} & {} \cr
           {\rm Along\ EF:} & {} & {\rm valid} & {} & {\rm breaks} \cr }$$} }
\end{figure}

\clearpage
\section{Concluding remarks}

\hs
To conclude, we have considered the propagation of open bosonic string, in presence of a
constant two-form $b$-field. Under appropriate Dirichlet boundary conditions, the string
dynamics reduces to that of a $D(3)$-brane in a constant EM-field. Incorporating
different orientations of a $B$-field with respect to the $E$-field, we have
performed an analysis, to explain some of the phenomena associated with the new light string modes
in a gravity decoupled theory. We emphasized on the presence of space and time-like windings
in the ($3+1$)-dimensional theory and explained the fate of Lorentzian signature of space-time there.
Taking account for a inherent relation between the space and time-like windings,
we have obtained bounds on the winding radii. It is shown 
that the time-like coordinate is bounded by the string scale and
describes a small dimension in the theory. As a result, the space coordinates in the  theory 
are argued to be large in comparison to the time dimension. The small time-like coordinate
implies discrete notion of time in the EM-string theory. 
For a large number of closely spaced time intervals, discrete notion may be
approximated to a continuous one. However, the continuous parameter does not
evolve with time and results in a time translation symmetry
in ($3+1$)-dimensions. It resolves the apparent paradox of "compact" time
in a Lorentzian signature of space-time. In other words, the notion of time breaks down in the
gravity decoupling limit and at the expense, temperature becomes significant there.
The phenomenon is supported by a 
non-trivial mixing of time with the space coordinates due to the additional winding energy
in the gauge-string spectrum. 

\sp
In addition, the presence of gauge field windings in the theory is emphasized and
it is argued that the the effective string scale is quantized in the units of "rigid" string
which may be interpreted as an electric dipole. 
Our analysis suggests that the effective string is made from a large
number of electric dipoles aligned along the direction of $E$-field.
In addition, the dipoles joined end to end in a series, since $T_{\rm eff}\ra 0$.
In other words,
a large space-like dimension gives rise to the "long" string description in the effective theory.
In fact the winding modes give
rise to nonlocality, which is an integral part of non-linearity in gauge field.

\sp
One of the important lesson, we learned, is that the EM-field has many interesting facets
in a open string theory. On the one hand, it describes the string dynamics on a noncommutative
space-time and on the other, it introduces the notion of thermal ensemble into a theory.
We have performed the thermal
string mode analysis to explicitly account for the winding modes in the theory. The mass-shell
condition for the thermal strings is analyzed and we have obtained an expression for the
critical (effective Hagedorn) temperature in the theory. The critical regime 
(${\cal T}\ra {\cal T}_H$), described by the area ABC in figure 2, is investigated
for transitions leading to the Hagedorn phase. Taking into account
the small/large dimensions, we have argued that the gauge-string phase is 
dominated by time-like
windings in the theory.

\sp
In the critical regime, we have obtained the corrections to (ordinary) $U(1)$ gauge
theory, due to its own non-linearity. The gauge-string corrections are due to the 
constant $b$-field in the theory. At ${\cal T}_H$, a spontaneous break down of 
noncommutative $U(1)$ symmetry is argued, analogus to that in QCD at its deconfinement
temperature. The symmetry 
breaking gives a mass to the $b$-field and forms condensate at ${\cal T}_H$.
In addition, we have shown the presence of tachyonic windings in the "long" string phase.
It is argued that a further increase in energy at ${\cal T}_H$, excites the string 
and decouples the tachyons from the phase. Then, the
Hagedorn phase is described by the ordinary (noninteracting)  gauge particles.
In the context, we have out-lined a generalization to noncommutative $U(N)$ theory,
in presence of $N$ number of coincident $D(3)$-branes. It will be interesting to analyze the
critical regime there, to work out the details of phase transitions leading to the Hagedorn phase.

\sp
In particular, the new windings, share some common properties between the finite temperature
\cite{fischler} and space-time noncommutative field and string theories. The fact that
the space and time-like windings naturally determine the large and small dimensions in a theory,
may shed some light on the low-scale gravity models. 

\sp
On the other hand, the gauge-string theory is a subject by itself and its study may provide
important tools to address some of the conceptual issues in quantum gravity with various 
alternate effective descriptions. 
For example, the genus zero $g=0$ contribution of gauge-string to the free-energy, possibly,
requires substantial correction to Riemannian manifold at short distances. It urges a 
non-perturbative
formulation and is beyond the scope of this paper.

\sp

\sp

\noindent
{\Large\bf Acknowledgments}

\sp
Author (S.K.) is grateful to S.D. Joglekar and A. Sen for useful
discussions, at the early stage of this work. He thanks P. Jain and K. Ray for
various discussions, in general, on the topic.
Some of the results obtained in this article,
were presented by him in Indo-Russian satellite
workshop on String theory and Integrable models, 22-23 January 2002 at Harish Chandra
Research Institute, Allahabad.

\vfil\eject

\def\anp{Ann. of Phys.}
\def\prl{Phys. Rev. Lett.}
\def\prd#1{{Phys. Rev.} {\bf D#1}}
\def\plb#1{{Phys. Lett.} {\bf B#1}}
\def\npb#1{{Nucl. Phys.} {\bf B#1}}
\def\mpl#1{{Mod. Phys. Lett} {\bf A#1}}
\def\ijmpa#1{{Int. J. Mod. Phys.} {\bf A#1}}
\def\rmp#1{{Rev. Mod. Phys.} {\bf 68#1}}



\begin{thebibliography}{99}

\baselineskip= 18 truept


\bibitem{douglas}M.R. Douglas, D. Kabat, P. Pouliot and S.H. Shenker, \npb{485} (1997) 85
({\tt hep-th/9608024}):

R. Dijkgraaf, E. Verlinde and H. Verlinde, \npb{500} (1997) 43
({\tt hep-th/9703030}):

A. Sen, Adv.Theor.Math.Phys.{\bf 2} (1998) 51 ({\tt hep-th/9709220}).


\bibitem{fradkin}E.S. Fradkin and A.A. Tseytlin, \plb{163} (1985)
123;

 A. Abouelsaood, C.G. Callan, L. Lovelace, C.R. Nappi, S.A. Yost,
\npb{280} (1987) 599;

C.P. Burgess, \npb{294} (1987) 427;

V.V. Nesterenko, \ijmpa {\bf 4} (1989) 2627.


\bibitem{ferrer}E.J. Ferrer, E.S. Fradkin and V. de la Incera, \plb{248} (1990) 281.

\bibitem{odintsov}A. Bytsenko, I. Lichtzier and S.D. Odintsov, Yad.Fiz.(Sov.J.Nucl.Phys.) 54,
(1991) 1453.

\bibitem{bachas}C. Bachas and M. Porrati, \plb{296} (1992) 77 ({\tt hep-th/9209032});

C. Bachas, \plb{374} (1996) 37 ({\tt hep-th/9511043}).

\bibitem{odintsov2}A. Bytsenko, S.D. Odintsov and L. Granda, \mpl{\bf A11} (1996) 2525.

\bibitem{gukov}S. Gukov, I.R. Klebanov and A.M. Polyakov, \plb{423} (1998) 64 
({\tt hep-th/9711112}).

\bibitem{connes}A. Connes, M.R. Douglas and A. Schwarz, JHEP
{\bf 2} (1998) 003 ({\tt hep-th/9711162});

M.R. Douglas and C. Hull, J.H.E.P. {\bf 9802} (1998) 008
({\tt hep-th/9711165}).

\bibitem{ambjorn}J. Ambjorn, Y.M. Makeenko, G.W. Semenoff and R.J. Sazbo,
({\tt hep-th/0012092}).


\bibitem{jabbari1}F. Ardalan, H. Arfaei, M.M. Sheikh-Jabbari, J.H.E.P. {\bf 9902} 
(1999) 016 ({\tt hep-th/9810072});

C.-S. Chu and P.-M. Ho, \npb{550} (1999) 151
({\tt hep-th/9812219});

V. Schomerus, JHEP {\bf 9906} (1999) 030
({\tt hep-th/9903205}).

\bibitem{kar1}S. Kar, \ijmpa{\bf 1} (2001) 41 ({\tt hep-th/9907117}).

\bibitem{seiberg}N. Seiberg and E. Witten, JHEP {\bf 9909} (1999) 032
({\tt hep-th/9908142}).

\bibitem{chu1}C.-S. Chu, P.-M. Ho and Y. -C. Kao, \prd{\bf 60} (1999) 126003
({\tt hep-th/9904133});

C.-S. Chu and P.-M. Ho, \npb{\bf 568} (2000) 447
({\tt hep-th/9906192});

M. Li and Y. -S. Wu, \prl{\bf 84} (2000) 2084
({\tt hep-th/9909085});

S. Kar, \npb{577} (2000) 171 ({\tt hep-th/9911251}); 
Int.J.Theor.Phys, Group Theory and Nonlinear Optics, in press ({\tt hep-th/0006073});

I. Cheplev and R. Roiban, JHEP {\bf 0103} (2001) 001
({\tt hep-th/0008090}).

\bibitem{minwalla}S. Minwalla, M. Van Raamsdonk and N. Seiberg, JHEP {\bf 0002}
(2000) 020 ({\tt hep-th/9912072}); 

M. Van Raamsdonk and N. Seiberg, JHEP {\bf 0003}
(2000) 035 ({\tt hep-th/0002186});

M. Hayakawa, {\tt hep-th/9912167};
{\tt hep-th/0009098}.

\bibitem{susskind}N. Seiberg, L. Susskind and N. Toumbas, JHEP
{\bf 0006} (2000) 021 ({\tt hep-th/0005040}). 

\bibitem{gopa1}R. Gopakumar, J. Maldacena, S. Minwalla and A. Strominger,
JHEP {\bf 0006} (2000) 036 ({\tt hep-th/0005048}).

\bibitem{gomis}J. Gomis and T. Mehen, \npb{\bf 591} (2000) 265
({\tt hep-th/0005129}).

\bibitem{klebanov}I. Klebanov and J. Maldacena, 
Adv.Theor.Math Phys.{\bf 4} (2000) 283 ({\tt hep-th/0006085}).

\bibitem{Sahakian}V. Sahakian, JHEP {\bf 0009} (2000) 025
({\tt hep-th/0008073}). 

\bibitem{alvarez1}L. Alvarez-Gaume and J.L.F. Barbon, \ijmpa{\bf 16} (2001) 1123 
({\tt hep-th/0006209}).

\bibitem{barbon1}J.L.F. Barbon and E. Rabinovici, \plb{\bf 486} (2000) 202
({\tt hep-th/0005073});

J.G. Russo and M.M. Sheikh-Jabbari, J.H.E.P.{\bf 07} (2000)
52 ({\tt hep-th/0006202}).

\bibitem{polyakov}A. Polyakov, \plb{72} (1978) 477.  

\bibitem{susskind2}L. Susskind, \prd{20} (1979) 2610. 

\bibitem{gross}D.J. Gross, \prl{\bf 60} (1988) 1229.

\bibitem{sathia}B. Sathiapalan, \prd{35} (1987) 3277;

Y.I. Kogan, JETP Lett. {\bf 45} (1987) 709;

K.H. O'Brien and C.-I. Tan, \prd{36} (1987) 1184.

\bibitem{atick}J.J. Atick and E. Witten, \npb{310} (1998) 291.

\bibitem{mukund}S.S. Gubser, S. Gukov, I.R. Klebanov, M. Rangamani and E. Witten,
J.Math.Phys.42 (2001) 2749 ({\tt hep-th/0009140}).

\bibitem{barbon2} J.L.F. Barbon and E. Rabinovici, JHEP {\bf 0106} (2001) 029
({\tt hep-th/0104169}). 

\bibitem{abel}S.A. Abel, J.L.F. Barbon, I.I. Kogan and E. Rabinovici, JHEP {\bf 9904}
(1999) 015;

M.C.B. Abdalla, A.L. Gadelha and I.V. Vancea, \prd{\bf64} (2001) 086005 
({\tt hep-th/0104068}). 

\bibitem{yoneya}T. Yoneya, Prog. in Theor.Phys.{\bf 103} (2000) 1081 ({\tt hep-th/0004024})

\bibitem{campbell}B.A. Campbell and K. Kaminsky, {\tt hep-th/0009098};

F.J. Petriello, \npb{\bf601} (2001)
169 ({\tt hep-th/01011109}).

\bibitem{witten}E. Witten, \npb{460} (1996) 335 ({\tt hep-th/9510135}).

\bibitem{aharony}O. Aharony, J. Gomis and T. Mehen, JHEP {\bf 0009} 
(2000) 023 ({\tt hep-th/0006236});

R. Cai and N. Ohta, JHEP {\bf 0010} (2000) 036
({\tt hep-th/0008119}).

\bibitem{fischler}W. Fischler, E. Gorbatov, A. Kashani-Poor, S. Paban and P. Pouliot,
JHEP {\bf 0005} (2000) 024 ({\tt hep-th/0002067}); 

G. Arcioni, J.L.F. Barbon, J. Gomis, M.A. Vazquez-Mazo, JHEP {\bf 0006}
(2000) 038 ({\tt hep-th/0004080}); 

Y. Kiem, D.H. Park and H.-T. Sato, \plb{499} (2001) 321 ({\tt hep-th/0011119});

L. Alvarez-Gaume, J.L.F. Barbon, R. Zwicky, JHEP {\bf 0105} (2001)
057 ({\tt hep-th/0103069}); 

C.-S Chu, J.Lukierski and W. Zakrzewski, {\tt hep-th/0201144};

F.T. Brandt, A. Das, J. Frenkel, D.G.C. McKeon and 
J.C. Taylor, {\tt hep-th/0204192}.  


\bibitem{kar3}S. Kar and Y. Kazama, \ijmpa{\bf 14} (1999) 1531
({\tt hep-th/9807239});

S. Kar, \npb{\bf 554} (1999) 163  ({\tt hep-th/9812230}).

\end{thebibliography}
\end{document}